\documentclass[twocolumn]{aastex631}

\usepackage{longtable}
\usepackage{threeparttable}  

\usepackage[figuresright]{rotating}
\usepackage{graphicx}
\usepackage{color}

\usepackage{tablefootnote}

\newcommand{\jwst}{\textit{JWST}}
\newcommand{\spitzer}{\textit{Spitzer}}
\newcommand{\herschel}{\textit{Herschel}}
\newcommand{\hst}{\textit{HST}}
\newcommand{\htwo}{H$_2$}

\newcommand{\feii}{[\ion{Fe}{2}]}

\newcommand{\neiii}{[\ion{Ne}{3}]}

\newcommand{\si}{[\ion{S}{1}]}

\newcommand{\ariii}{[\ion{Ar}{3}]}
\newcommand{\asec}{$^{\prime\prime}$}

\newcommand{\kms}{km\,s$^{-1}$}
\newcommand{\cmt}{cm$^{-3}$}
\newcommand{\cmd}{cm$^{-2}$}

\newcommand{\um}{$\mu$m}

\newcommand{\msunyr}{M$_{\odot}$\,yr$^{-1}$}

\begin{document}

\title{PROJECT-J: the shocking \htwo\ outflow from HH46} 

\correspondingauthor{Maria Gabriela Navarro}
\email{maria.navarro@inaf.it}

\author[0000-0002-1860-2304]{Maria Gabriela Navarro}
\affiliation{INAF - Osservatorio Astronomico di Roma, 
Via di Frascati 33, 00078 Monte Porzio Catone, Italy}

\author[0000-0002-9190-0113]{Brunella Nisini}
\affiliation{INAF - Osservatorio Astronomico di Roma, 
Via di Frascati 33, 00078 Monte Porzio Catone, Italy}

\author{Teresa Giannini}
\affiliation{INAF - Osservatorio Astronomico di Roma, 
Via di Frascati 33, 00078 Monte Porzio Catone, Italy}

\author{Patrick, J. Kavanagh}
\affiliation{Department of Physics, Maynooth University, Maynooth, Co. Kildare, Ireland}

\author[0000-0001-8876-6614]{Alessio Caratti o Garatti}
\affiliation{INAF - Osservatorio Astronomico di Capodimonte, Salita Moiariello 16, 80131 Napoli, Italy}

\author{Simone Antoniucci}
\affiliation{INAF - Osservatorio Astronomico di Roma, 
Via di Frascati 33, 00078 Monte Porzio Catone, Italy}

\author{Hector G. Arce}
\affiliation{Department of Astronomy, Yale University, New Haven, CT 06511, USA}

\author[0000-0001-5776-9476]{Francesca Bacciotti}
\affiliation{INAF - Osservatorio Astrofisico di Arcetri, 
Largo E. Fermi 5, I-50125 Firenze, Italy}

\author[0000-0002-1593-3693]{Sylvie Cabrit}
\affiliation{LERMA, Observatoire de Paris-PSL, Sorbonne Université, CNRS,  F-75014 Paris, France}
\affiliation{IPAG, Observatoire de Grenoble, Université Grenoble-Alpes, France}

\author[0000-0002-2210-202X]{Deirdre Coffey}
\affiliation{University College Dublin, School of Physics, Belfield, Dublin 4, Ireland}

\author{Catherine Dougados}
\affiliation{IPAG, Observatoire de Grenoble, Université Grenoble-Alpes, France}

\author[0000-0001-6496-0252]{Jochen Eisl\"offel}
\affiliation{Thüringer Landessternwarte, Sternwarte 5, D-07778 Tautenburg, Germany}

\author[0000-0002-5380-549X]{Patrick Hartigan}
\affiliation{Physics and Astronomy Dept., Rice University, 
6100 S. Main, Houston, TX 77005-1892, USA}

\author[0000-0002-6296-8960]{Alberto Noriega-Crespo}
\affiliation{Space Telescope Science Institute, 3700 San Martin Drive, Baltimore, MD, 21218, USA}

\author{Linda Podio}
\affiliation{INAF - Osservatorio Astrofisico di Arcetri, Largo E. Fermi 5, I-50125 Firenze, Italy}


\author{Ewine F. van Dishoeck}
\affiliation{Leiden Observatory, Leiden University, PO Box 9513, NL 2300, RA Leiden, The Netherlands}

\author[0000-0002-3741-9353]{Emma T. Whelan}
\affiliation{Department of Physics, Maynooth University, Maynooth, Co. Kildare, Ireland}



\begin{abstract}

We analyze the \htwo\ emission observed in the HH46 Class I system as part of PROJECT-J (PROtostellar JEts Cradle Tested with \jwst), to investigate the origin and excitation of the warm molecular outflow.
We used NIRSpec and MIRI spectral maps (1.6–27.9 \um) to trace the structure and physical conditions of the outflow. By fitting the \htwo\ rotational diagrams with a multi-temperature gas model, we derived key physical parameters  including temperature, extinction, column densities, and the ortho-to-para ratio. This information is combined with a detailed kinematical analysis and comparison with irradiated shock models.
We find no evidence of \htwo\ temperature or velocity stratification from the axis to the edge of the outflow, as would be expected in MHD disk-wind models and as observed in other outflows. Instead, the observations suggest that the \htwo\ emission arises from shock interactions between jet bow shocks and/or wide-angle winds with the ambient medium and cavity walls. NIRSpec emission and velocity maps reveal expanding molecular shells, likely driven by the less luminous source in the binary system. 
We infer an accretion rate of $\la$ 10$^{-9}$ \msunyr for the secondary source, approximately one order of magnitude lower than that of the primary.
The \htwo\ emission is consistent with excitation by low-velocity ($\sim10$ \kms) J-type shocks, irradiated by an external UV field that may originate from strong dissociative shocks driven by the atomic jet. Future \jwst\ observations will further constrain the evolution of the expanding shell and the mechanisms driving the outflow.

\end{abstract}

\section{Introduction} \label{sec:intro}
Investigating protostars and their accompanying planet-forming disks requires a profound understanding of the associated outflows that shape their evolution. 
Within protostellar systems—comprising young stellar objects, compact accretion disks, and dusty envelopes—mass ejection through powerful jets and winds is a fundamental process. These outflows, revealed by their strong line emission and extending from au to parsec scales, play a key role in stellar evolution by removing mass and angular momentum from the forming system \citep{ray2021,bally2016,frank2014}.

Current understanding suggests that jets are magneto-centrifugally launched from the star-disk interaction regions. In magneto-hydrodynamic (MHD) disk-wind models \citep[e.g.][]{pelletier1992, ferreira1997, bai2016}, high-velocity collimated jets, reaching speeds of 200-400 \kms, originate from the inner disk regions. Furthermore, slower and less collimated flows, with velocities ranging from 5 to 30 \kms, are launched from extended radii over the disk surface, profoundly influencing disk physics and the formation of planetary systems \citep{lesur2023, pascucci2023}. On the other hand, X-wind models \citep{shu2000,shang2020}, propose a small region at the disk's inner edge as the origin of outflows, that propagates in a wide fan of streamlines. Finally, additional models consider that the outflow is directly launched from the stellar magnetosphere, powered by the accretion energy released on the stellar surface \citep[][e.g.]{Matt&pudritz2005, zanni2013,romanova2009}. 
The distribution of the gas excitation in these models results in the appearance of a collimated axial high velocity jet surrounded by a molecular wide angle wind. 

The launch and propagation of jets and winds intricately influence the ambient medium, forming large cavities in infalling envelopes and entrainment of cold ambient gas. Deciphering these mechanisms requires scrutiny into regions within a few hundred au from the central star, where outflows retain crucial information about their velocity, collimation, and connection with accretion events. To achieve this goal, high angular resolution observations, i.e. $<$ 100 au,  are required.

Studies aimed at investigating the jet and wind launching mechanisms have predominantly focused on relatively evolved pre–main-sequence stars \citep[Classical T Tauri or Class II stars, ages of $\sim$10$^6$–10$^7$ yr; e.g.][]{pascucci2023}. However, it is equally crucial to study younger systems in the earlier phases of their evolution—namely the Class 0/I stage (ages $<$ 10$^6$ yr)—when vigorous accretion is still ongoing. These sources remain deeply embedded in dusty envelopes, which obscure their inner warm and active regions at optical wavelengths due to the high line-of-sight extinction. Consequently, current observational constraints on the jet and wind origin in these objects remain inconclusive \citep[e.g.][]{nisini2016,nisini2015,watson2016}.

In this context, the James Webb Space Telescope \citep[\jwst,][]{Gardner2023}
has revolutionized our ability to probe the earliest stages of jet/outflow formation and their interaction with the ambient medium owing to its high angular resolution and spectral coverage at infrared wavelengths. 
Leveraging \jwst's capabilities, several observations are being conducted to comprehensively study protostellar outflows, offering unprecedented insights into their morphology, composition, and dynamics \citep[e.g.][]{yang2022,harsono2023, narang2024,narang2025,barsony2024,assani2024,caratti2024,delabrosse2024,federman2024,Tychoniec2024,legouellec2025,vandishoeck2025,Vleugels25}. 

One of the most complete observational programs in terms of sensitivity, spectral range (extending down to 1.6 \um), and spatial coverage is the PROtostellar JEts Cradle Tested with \jwst\ (PROJECT-J).
The aim is to achieve a deep understanding of one of the most iconic examples of Class I Young Stellar Object (YSO), namely HH46 IRS and its associated HH46/47 objects, through MIRI Medium Resolution Spectrograph (MRS) and NIRSpec Integral Field Spectroscopy (IFU) observations mapping the source and the base of its outflow \citep[][hereafter Paper I]{nisini2024}.

HH46 IRS is an embedded (A$_V>$ 35 mag) binary YSO \citep[separation $\sim$0.\asec2;][Paper I]{reipurth2000}, located in the Gum Nebula at a distance of 450 pc. Throughout this paper, we refer to the primary and secondary components as source a$^\star$ and source b$^\star$, respectively.
The object drives the well-studied Herbig-Haro parsec scale outflow HH46/47, with an estimated dynamical age of
$<$ 10$^5$ yr \citep{stanke1999}. Its visible blue-shifted part has been studied at length at optical wavelengths \citep[e.g.][]{eisloffelpm1994,heathcote1996,hartigan2011}. The high-velocity ($v\geq300$ \kms) blue-shifted jet presents a prominent wiggling, attributed to the presence of a tertiary, unresolved, companion \citep{reipurth2000}.
The red-shifted outflow has been instead observed at near-IR wavelengths, but only relatively far from the central object due to the significant extinction at its base \citep{eisloffelh21994,reipurth2000,jochen2000,erkal2021}. Space mid- and far-IR observations have been able to peer closer to the HH46 IRS source,  although at low spatial resolution and with considerable contamination from the continuum emission from the central source (\spitzer: \citealt{noriega2004}, \herschel: \citealt{nisini2015}, \hst: \citealt{erkal2021}). The HH46/47 system is associated with a bipolar wide-angle molecular outflow studied at mm and sub-mm wavelengths \citep{vankempen2009,arce2013}. 
The most recent high-resolution ALMA observations show the presence of a conical cavity, within which multiple shell structures at different velocities are detected, suggesting the presence of bow shocks. 
This configuration has been interpreted as due to the entrainment of the ambient medium by a variable wide-angle wind caused by accretion-driven episodic ejections \citep{zhang2016,zang2019} or by a succession of large bow-shocks driven by a variable narrow jet \citep{Rabenanahary2022}. 

\begin{figure*}[ht]
    \centering
\includegraphics[width=1\textwidth,keepaspectratio]{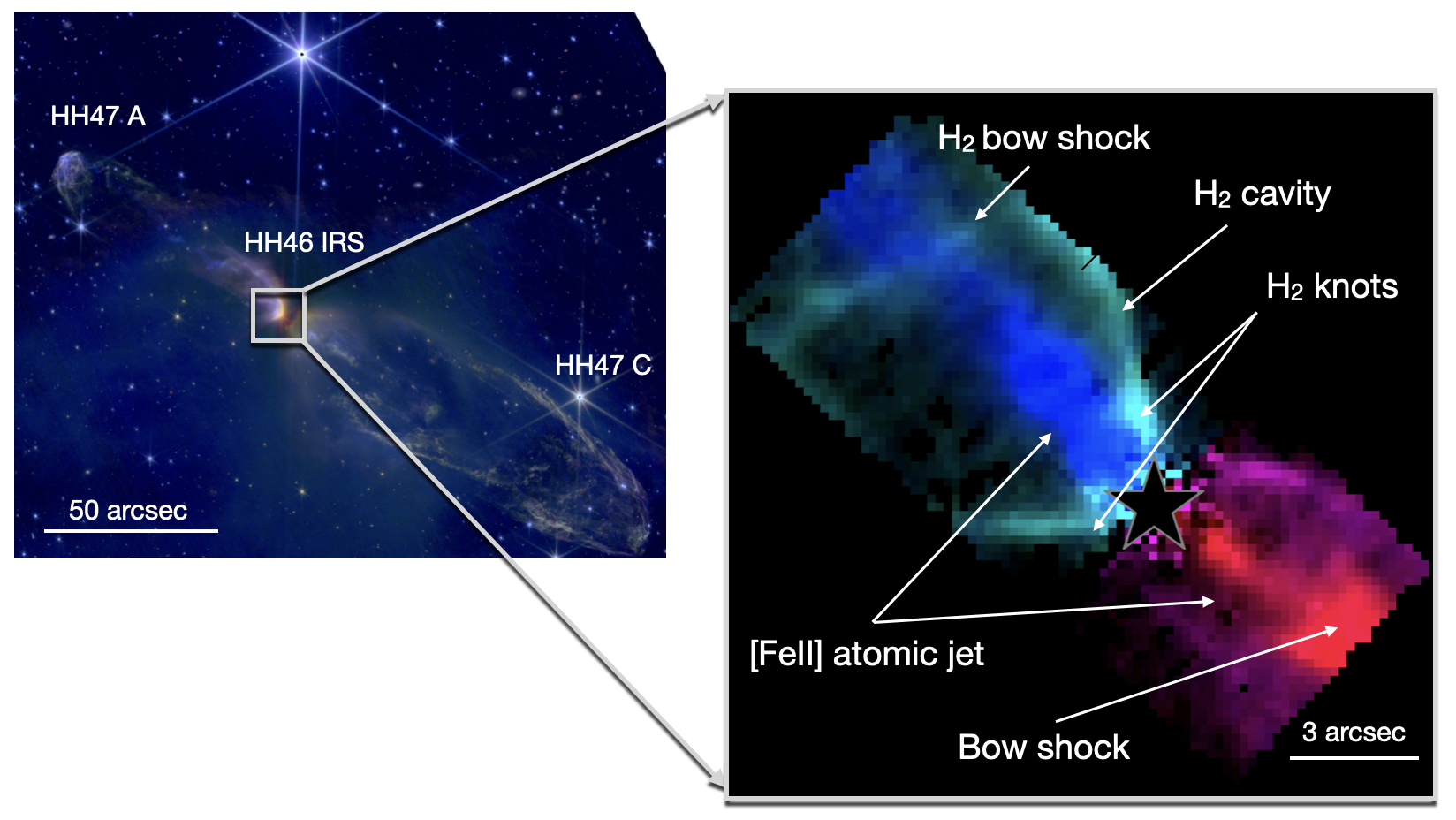}
    \caption{Left:
    \jwst/NIRCam image of HH~46/47, taken from the Mikulski Archive for Space Telescopes (MAST), obtained as part of the DDT program PID4441 (P.I. K. Pontoppidan). The image combines the F200W and F335M filters, which primarily trace the \htwo\ emission, while the F115W filter was used as a jet tracer. 
    Right: Enlargement of the region observed by PROJECT-J. The \htwo\ emission from the red- and blue-shifted lobes is shown in magenta and cyan, respectively. The \feii\ 5.5 \um\ emission tracing the red- and blue-shifted atomic jet is overlaid in red and blue, respectively. The black star marks the source position. The main structures identified in Paper I—such as the atomic jet, molecular cavity, compact knots and bow shocks—are labeled for reference.
    }
    \label{fig:nircam}
\end{figure*}
The first results of PROJECT-J have been presented in Paper I (see Figure \ref{fig:nircam}). The observations highlighted the complex HH46 outflow morphology, where the inner jet, detected in multiple bright forbidden lines, appears highly misaligned with respect to the axis of the blue-shifted molecular cavity and the molecular shock interaction regions. 
The latter are traced by \htwo\ emission appearing as compact knots and large bow shocks located at different distances from the central source.

The red-shifted part of the outflow, observed for the first time within $\sim$ 2000 au from the source, presents instead a more symmetric morphology, with the atomic jet pushing a large bow shock observed in \htwo\ emission. In Paper I, it is suggested that the significant asymmetries observed between the blue-shifted atomic jet and the \htwo\ emission, and between the outflow lobes, could be due to the influence of two outflows being driven by each of the components of the HH46 IRS binary system, which has been resolved by \jwst/NIRCam observations. The interaction with an inhomogeneous environment can further contribute to the observed morphological and kinematical asymmetries.
Bright \htwo\ is also observed throughout the walls of the conical cavity opened by the outflow, nested within the CO cavity observed with ALMA. 

The emerging picture highlights a complex interplay between the collimated jet, multiple nested molecular shells, and bright asymmetric outflows. In this work, we present a detailed analysis of the \htwo\ emission along the HH 46 outflow to disentangle the underlying physical and kinematical components.
Previous \htwo\ observations of the HH46 system were limited to low spatial resolution imaging of the \htwo\ 2.12 \um\ line \cite{eisloffelh21994} and long-slit near-IR spectroscopy along the jet \cite{garcialopez2010}. More recently, \cite{birney2024} have studied the \htwo\ kinematics in the inner blue-shifted region through VLT/SINFONI observations.

The exquisite quality of the PROJECT-J observations, mapping \htwo\ pure rotational and ro-vibrational lines with excitation energies from a few hundred K up to more than 20000 K, allows us to investigate the mechanisms of dynamics, excitation and heating of the outflows, thus providing an important piece of information in the study of  the origin of the molecular wind and of its connection the collimated atomic jet. 


The paper is organized as follows. Section~\ref{sec:red} describes the observations and data reduction procedures. Section~\ref{sec:Results} presents the main results, including continuum-subtracted line maps and spectra extracted from representative regions. Section~\ref{sec:Analysis} provides a detailed analysis of the \htwo\ emission through rotational diagrams constructed from MIRI and NIRSpec data, from which physical parameters such as temperature stratification, extinction, column density, and velocity maps are derived. A comparison with shock models for selected regions is also presented. Section~\ref{sec:disc} discusses the physical conditions and origin of the \htwo\ emission structures, and includes an estimate of the mass-loss rate. Finally, Section~\ref{sec:concl} summarizes the main conclusions of this work.

\section{Observations and Data reduction} \label{sec:red}
The observations analysed here are part of the \jwst\ Cycle 1 program PID1706 (P.I. B. Nisini), conducted using the Mid-InfraRed \citep[MIRI,][]{rieke2015, wright2023} and NIRSpec Instruments \citep[NIRSpec,][]{Jakobsen2022} in February and March of 2023. Both instruments employed the Integral Field Unit (IFU) modes (MIRI-MRS and NIRSpec-IFU).

The detailed description of the observation setup and steps for the data reduction are given in Section 2 of Paper I. In summary, MIRI observations were executed with three grating settings (SHORT, MEDIUM, LONG) and four channels (1, 2, 3, 4), encompassing the entire available wavelength range covering 4.9–27.9 \um.
The MIRI-MRS pixel scale increases with wavelength (from 0.\asec196 at 5 \um\ to 0.\asec245 at 17 \um), while the field of view expands from 3.\asec2$\times$3.\asec7 to 5.\asec2$\times$6.\asec2. The diffraction-limited spatial resolution also increases with wavelength, from $\sim$0.\asec2 at 5 \um\ to $\sim$0.\asec6 at 17 \um\ \citep{Patapis24, Law25}.
To cover a broader area, 4×2 mosaics were employed, resulting in a total coverage of 6\asec $\times$15\asec\ at the shortest wavelength and $\sim$ 11\asec $\times$20\asec\ at the longest, with a nominal resolving power ranging from $\sim$3 710 at 5 \um\ to $\sim$1 330 at 28 \um\ (i.e. $\Delta v \approx 80$--$225~\mathrm{km\,s^{-1}}$; \citealt{Wells15}; \citealt{Argyriou2023}; JWST Documentation).
The mapped region covers the central object, the atomic jet, and the base of the \htwo\ molecular cavity and shocks, as shown in Figure \ref{fig:nircam}.

NIRSpec observations were executed with two distinct grating settings, G235H and G395H, with a nominal resolving power of $\sim$2 700 (i.e. $\Delta v \sim 110$ \kms), spanning a total wavelength range of 1.66–5.27 \um. 2×2 mosaics were employed, covering a region of 6\asec $\times$6\asec (see Figure 1 of Paper I for a map of the covered region). 

A subsequent reduction was performed using version 1.14.0 of the pipeline and its corresponding Calibration Reference Data System (CRDS) context \verb|jwst_1236.pmap|.
In addition, multiple corrections were applied at various stages of the reduction process to mitigate artifacts and enhance overall data quality. 
Some of the most significant adjustments include the identification of warm pixels in both instruments, the implementation of astrometric corrections conducted during simultaneous MIRI imaging observations, background subtraction outside the pipeline for MIRI, and the application of residual dark stripping correction to address the vertical pattern observed in NIRSpec data (see Section 2 of Paper I). 

Compared to the CRDS context used to reduce the data presented in Paper I (version 1.11.1, CRDS context \verb|jwst_1094.pmap|),  improvements have been implemented in the pipeline, reducing flux calibration errors. For MIRI, the nominal uncertainty is now 1-2$\%$ \citep{Law25}.
This value is consistent with the estimated intercalibration uncertainty, as inferred from the discrepancies observed between the subchannels of the source spectrum.
The intercalibration uncertainty between NIRSpec and MIRI for the source spectrum is quantified to be approximately 5$\% $.
To further validate the flux calibration, we compared the measured flux of the \htwo\ 0-0 S(8) line at 5.05 \um\ between the two instruments. The fluxes were found to differ by $<$5$\%$, in agreement with the expected calibration accuracy.

\section{Results} \label{sec:Results}
\subsection{Continuum subtracted MIRI and NIRSpec cubes} \label{sec:contsub}
Obtaining accurate \htwo\ line images of the HH46 outflow requires proper subtraction of the continuum emission from both the central source and the surrounding nebulosity, which dominates especially in the mid-IR range. 

 We performed a linear fit of the local continuum around each line of interest ($\Delta\lambda\sim0.1$ \um), masking both the line and, when necessary, other nearby emission lines that may appear close to the region of interest in the spectrum. Subsequently, the linear fit is subtracted from the spectrum. This procedure is performed in each cube pixel, resulting in a sub-cube centered around the line.

\citet{birney2024} demonstrate that part of the 2 \um\ line emission observed within the HH46 cavity may result from scattered light originating near the central source.
They performed a removal of this scattered light contribution by extracting a spectrum of the HH46 IRS central source and subtracting it at each pixel of their SINFONI map, after rescaling. We do not perform such a correction to avoid removing in the process also some compact emission from the outflow present in the source extraction region.
However, from the \cite{birney2024} analysis we see that the scattered contribution at 2 \um\ is relevant only for the diffuse \htwo\ emission seen in the cavity.
Moreover, we expect the impact of scattering to decrease with wavelength, following a $\lambda^{-4}$ dependence, although the presence of large dust grains could sustain a non-negligible scattering contribution up to 10 \um\ (e.g., \citealt{tazaki2025}).




\subsection{Line intensity maps} \label{sec:linemap}
The continuum-subtracted cubes obtained as described in Section \ref{sec:contsub} have been used to create line intensity maps of representative \htwo\ lines at different excitation conditions. As described in Paper I, MIRI observations detect only the \htwo\ pure-rotational lines from the v=0 vibrational level, from S(1) to S(8), corresponding to upper-level excitation energies between E$_{up}\sim$ 1015 - 8677 K. In contrast, the NIRSpec spectra reveal not only 0–0 transitions up to S(15), but also numerous ro-vibrational transitions from the v=1 and v=2 levels, with excitation energies reaching up to E$_{up}\sim$30000 K. Transitions from the v=3 level are observed only in a few compact, high-emission knots.

Line intensity maps are obtained by integrating the spectral elements that cover the considered line profile. These maps highlight the molecular outflow morphology and how its structure changes for different excitation regimes. 

\begin{figure}[t]
    \centering
\includegraphics[width=0.5\textwidth,keepaspectratio]{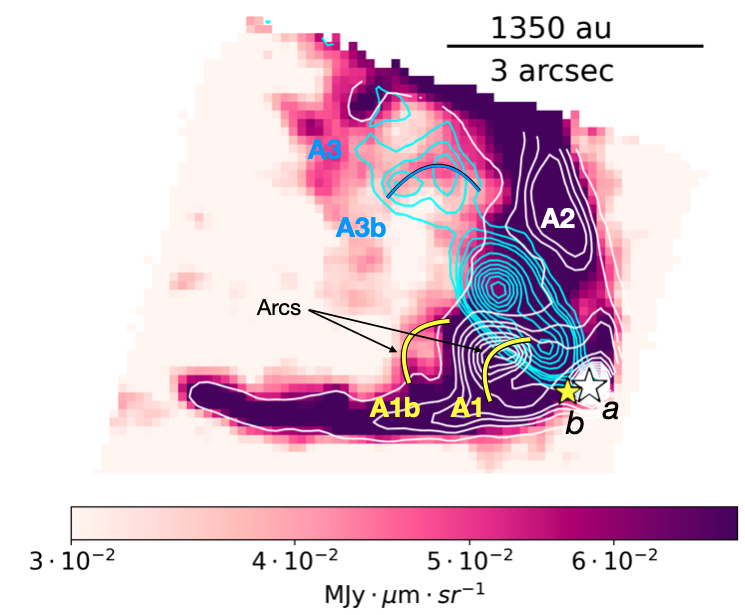}
    \caption{
\jwst/NIRSpec stacked \htwo\ line intensity map, obtained by combining the $v=1$–0 ro-vibrational transitions from S(1)–S(3) and O(2)–O(7). White contours delineate the brightest regions of the stacked \htwo\ emission (levels from 0.085 to 2.0~MJy~sr$^{-1}$), highlighting the arc-shaped morphology of knot A1. 
Cyan contours trace the \feii\ 1.81~\um\ emission (levels from 0.028 to 0.4~MJy~sr$^{-1}$), outlining the collimated jet. 
The two components of the binary system are marked with stars: the white star denotes the primary source (a$^\star$), and the yellow star marks the secondary (b$^\star$). 
Blue and yellow lines indicate the newly identified A3b and A1b arcs, associated with the jet from a$^\star$ and the outflow from b$^\star$, respectively.}
    \label{fig:nirspec_linemap}

\end{figure}

Although the NIRSpec map covers a smaller area than the MIRI observations — limited to the inner portion of the blue-shifted outflow — its higher spatial resolution enables a more detailed analysis of the underlying structures.
A first inspection of the intensity maps of the brightest ro-vibrational lines covered by the NIRSpec cube reveals no significant morphological differences compared to the \htwo\ 2.12 \um\ map presented in Paper I, regardless of the excitation energy of the lines.
Therefore, we have stacked together maps of several bright lines to obtain a higher SNR image of the \htwo\ ro-vibrational emission. We considered in particular the v=1-0 ro-vibrational lines from S(1)-S(3) and O(2)-O(7), which are those with the higher SNR.
Figure \ref{fig:nirspec_linemap} shows this stacked \htwo\ intensity map. 

The \htwo\ emission follows the borders of the paraboloid cavity and we have shown in Paper I that the edges of the \htwo\ emission lay inside the dust scattered nebula as traced by the 2\um\ continuum. In addition to the cavity walls, the brightest \htwo\ emission comes from the two knots (identified as A1 and A2), and from a bow shock (A3).
However, additional weaker curved structures are also detected in this region.
In particular, we identify one 
additional emission bow shock located farther out from the A1 knot, designated as A1b. 
The white contours allow us to distinguish the shape of the A1 knot, which appears rounded, suggesting the presence of an expanding shell or a bow shock. 
Additionally, another arc, A3b, is detected in the northern region, internal to the A3 arc.
The spatial coincidence of these \htwo\ bow shocks with peaks of \feii\ emission provides evidence of the interaction between A3 and A3b with the collimated jet. Overall, none of the high excitation lines is tracing the jet.

\begin{figure*}[ht]
    \centering
    \includegraphics[width=0.95\textwidth,keepaspectratio]{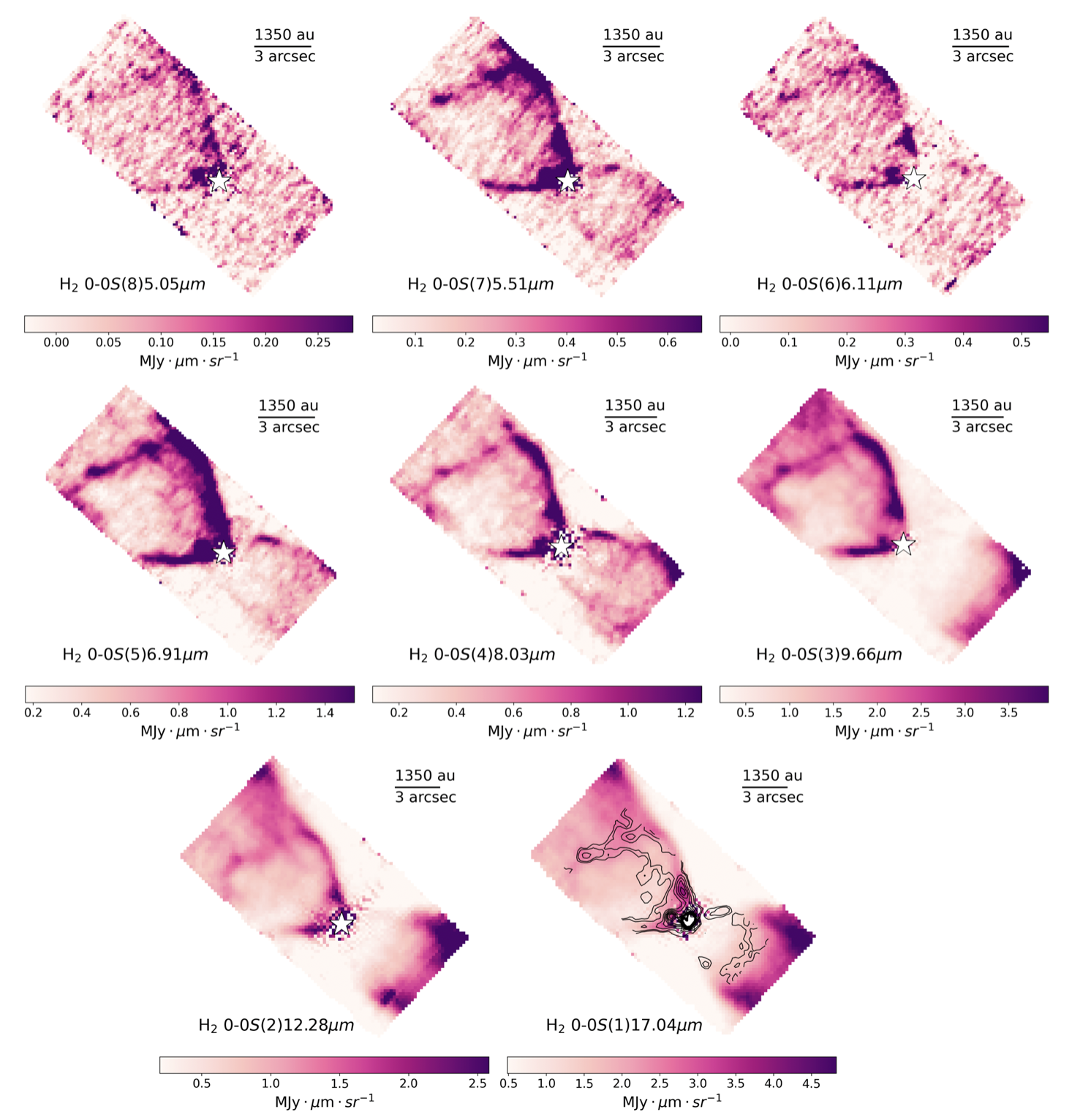}
    \caption{MIRI continuum-subtracted \htwo\ line intensity maps of pure rotational transitions. White stars mark the positions of the two sources. Contours of the 0–0~S(7) line at 5.51~\um\ (levels from 0.6 to 10.0~MJy~sr$^{-1}$) are overlaid on the 0–0~S(1) map at 17.04~\um. 
The contours clearly show that the MIRI field of view varies with wavelength, ranging from 6\asec$\times$15\asec\ for the 0–0~S(8) line at 5.05~\um\ to 8\asec$\times$17\asec\ for the 0–0~S(1) line at 17.04~\um.}
    \label{fig:MIRI_image}
\end{figure*}

The morphology of the \htwo\ gas at lower excitation is shown in Figure \ref{fig:MIRI_image}, where MIRI maps of the pure rotational lines from S(1) to S(8) are displayed. The MIRI maps cover a much larger region, including the red-shifted outflow. 
The \htwo\ emission extends throughout the entire cavity, being particularly bright at the edges. The structures presented in Figure \ref{fig:nircam} and in Paper I (prominent emission peaks and bow shocks)
are present in all the various excited lines. The lines of higher excitation, tracing the warmer gas, exhibit the mentioned structures more distinctly. In contrast, at lower excitation, the emission of colder gas tends to become more diffuse.
The redshifted bow-shock is prevalently seen in the 12 \um\ and 17 \um\ lines, located just outside or at the edge of the FoV at the shorter wavelengths (channel 1).

The opening angle of the \htwo\ cavity was measured for all detected transitions, yielding values between 77° and 79°, with an uncertainty of $\pm3°$, consistent within the measurement errors. This quantitative result confirms that the morphology of the \htwo\ emission remains similar across excitation levels, with no evidence of increased collimation at higher energies.
Overall, the morphology of the lines at different excitation energies does not show macroscopic variations. In contrast to other outflows \citep[e.g.][Navarro et al., in prep.]{caratti2024,Tychoniec2024}, where a layered structure is observed—with lines at increasing upper energy levels exhibiting smaller opening angles relative to the outflow axis—here all transitions show comparable collimation. A similar behavior has also been reported for the CED110IRS4 outflow by \citet{narang2025}.

\subsection{Spectra of representative regions}

\begin{figure}[t]
    \centering
\includegraphics[width=0.5\textwidth,keepaspectratio]{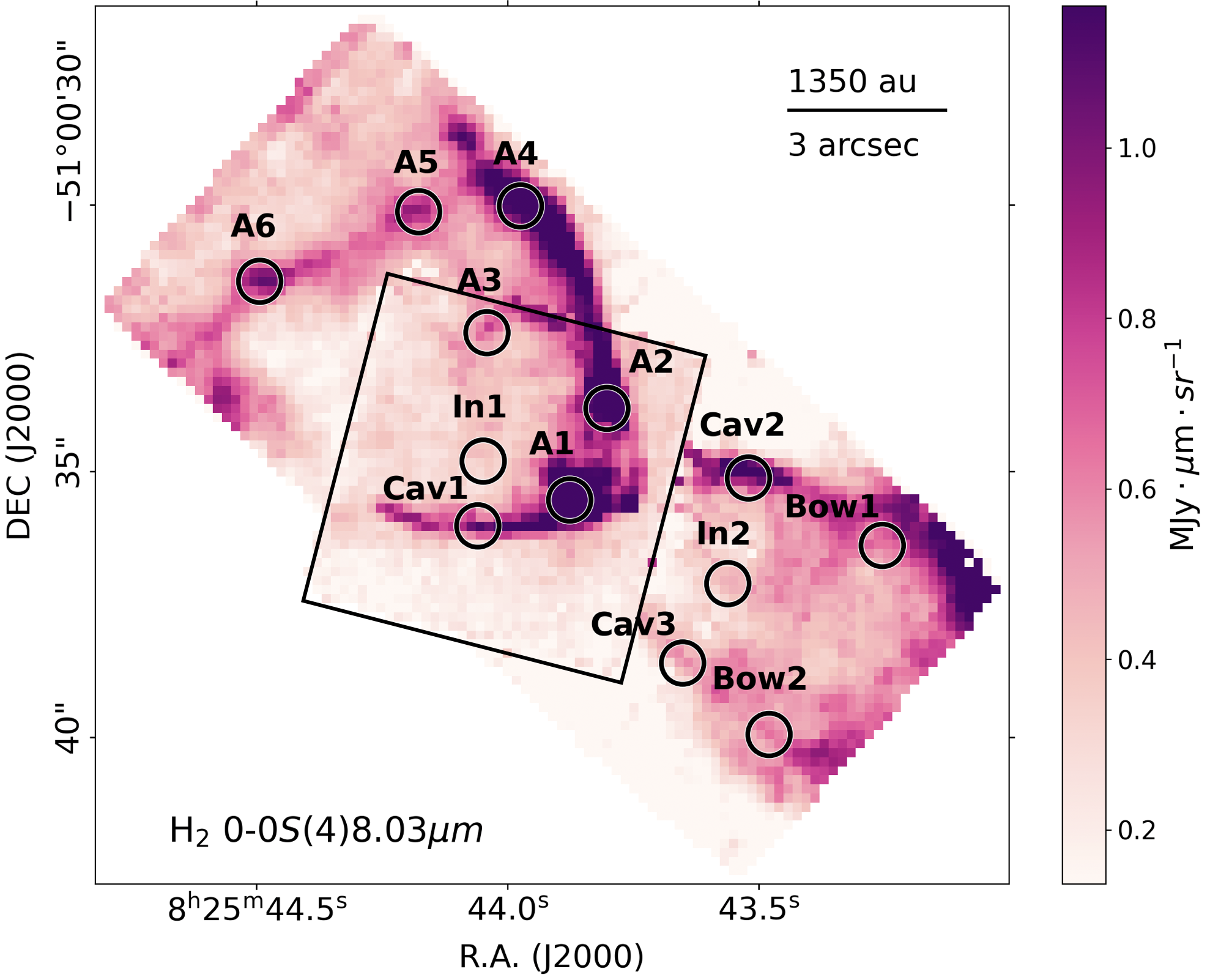}
\caption{Continuum-subtracted map of the \htwo\ 0–0~S(4) line at 8.03~\um. 
The NIRSpec field of view is outlined by a black square. 
Representative regions within the outflow, selected for further analysis, are indicated with white circles. 
Each circular aperture has a radius of 0.\asec4.}
    \label{fig:regions}
\end{figure}

We defined 13 representative regions of different observed features to analyze in detail variations of the \htwo\ excitation conditions in the outflow.
Figure \ref{fig:regions} displays the continuum-subtracted image of the \htwo\ 0–0 S(4) line at 8.03 \um\ along with the selected regions. 
We identified the prominent \htwo\ emission peaks A1 and A2 near the central source, as well as peak positions inside the extended bow shocks located at larger distance in the blueshifted outflow, reported in Paper I as A4, A5, and A6 arcs.
Moreover, we defined additional areas tracing emission from the cavity walls (Cav1, Cav2, and Cav3), the wings of the prominent redshifted bow shock (Bow1 and Bow2), and regions internal to the blue- and redshifted cavity (In1 and In2). The regions were selected based on the coverage across all MIRI channels, excluding the edges at shorter wavelengths where the field of view is smaller.
Table \ref{table:reg} lists the coordinates for the identified regions and briefly describes each of them.

\begin{table*}[ht]
\centering
\caption{Regions selected for the analysis (see Figure \ref{fig:regions})}
\begin{tabular}{c|c|c|c}
\hline
\hline
Region & R.A. & DEC & Description \\
\hline
 \multicolumn{4}{c}{Blue-shifted outflow}   \\ \hline
A1 & 8:25:43.8767 & -51:00:35.541 & Emission knot  \\ 
A2 & 8:25:43.8027 & -51:00:33.820 & Emission knot \\ 
A3 & 8:25:44.0414 & -51:00:32.400 & Emission arc \\ 
Cav1 & 8:25:44.0596 & -51:00:36.027 & Cavity edge\\
In1 & 8:25:44.0488 & -51:00:34.811 & Inner cavity region \\ 
A4 & 8:25:43.9747 & -51:00:30.020 & Knot along the northern cavity wall \\
A5 & 8:25:44.1773 & -51:00:30.128 & Emission peak in the N arc \\
A6 & 8:25:44.4937 & -51:00:31.434 & Emission peak in the NE arc \\ \hline
 \multicolumn{4}{c}{Red-shifted outflow}   \\ \hline
Cav2 & 8:25:43.5209 & -51:00:35.128 & Northern cavity edge\\
Cav3 & 8:25:43.6519 & -51:00:38.602 & Southern cavity edge\\
Bow1 & 8:25:43.2546 & -51:00:36.394 & Northern bow shock wing\\ 
Bow2 & 8:25:43.4800 & -51:00:39.944 & Southern bow shock wing\\
In2 & 8:25:43.5620 & -51:00:37.111 & Inner cavity region \\
\hline
\end{tabular}
\label{table:reg}
\end{table*}

Spectral extraction was carried out using a circular aperture with a radius of 0.\asec4, thus larger than the expected MIRI-MRS PSF FWHM. According to the JWST documentation, the FWHM rises from $\sim$ 0.\asec3 at 5 \um\ to $\sim$ 1.\asec0 at 25 \um, implying a value of order $\sim$0.\asec6 at 17 \um\ (\citealt{Argyriou2023}; JWST Documentation).
Due to the variation in pixel scale across MIRI sub-channels, the number of pixels encompassed by the aperture changes as a function of wavelength.

The full (MIRI+NIRSpec) extracted spectrum for the region A1 is presented in Figure 16 of Paper I. As mentioned in Section \ref{sec:linemap}, \htwo\ lines are detected up to the v=3-2 ro-vibrational level. Figures \ref{fig:NIRSPEC_specs1}, \ref{fig:NIRSPEC_specs2} and \ref{fig:MIRI_specs} in the Appendix present the extracted spectra of the selected regions covering the NIRSpec G235H, G395H and MIRI FoV, respectively. 

Fluxes of the detected transitions were derived by integrating the emission lines corresponding to the best-fit Gaussian profiles. The resulting values are reported in Tables \ref{tab:nirspec_lines} and \ref{tab:miri_lines} of the Appendix, corresponding to NIRSpec and MIRI observations, respectively, for the regions defined in Table \ref{table:reg}. The reported errors are estimated from the local RMS around each line. Lines with SNR$<$3 were ignored.

\section{Analysis} \label{sec:Analysis}

\subsection{Rotational diagrams}

To estimate the physical conditions of the \htwo\ emitting gas (specifically temperature and column density) and the visual extinction, rotational diagrams for the regions observed with MIRI and NIRSpec have been constructed. In these diagrams, the natural logarithm of the \htwo\ column density in the upper rotational state ($N_u$), inferred from the observed line intensities, divided by the degeneracy of that state ($g_u$) is plotted against the upper state energy (E$_{up}$), where $g_u= (2j+1)(2s+1)$.

For these diagrams we considered only lines with SNR$>$5. These diagrams were analyzed by applying two different approaches, namely a two-components fit and a temperature stratification fit, which are separately described in the following sections. 

\subsubsection{Bimodal temperature distribution}\label{sec:dfit}

In the assumption that the \htwo\ gas is in Local Thermodynamical Equilibrium (LTE) at a single temperature and with the equilibrium ortho-to-para ratio (OPR) of 3, we would expect the data points to form a straight line, with the inverse of the slope providing the \htwo\ rotational temperature ($T_{\rm H_2}$) and the intercept indicating the total \htwo\ column density ($N_{\rm H_2}$).

The visual extinction (A$_V$) was determined before performing the temperature fitting and used to correct the observed line fluxes for reddening. The resulting extinction-corrected fluxes were then used to derive the column densities, under the assumption of optically thin emission.

A$_V$ has been estimated by combining two independent methods.
1) we used the \htwo\ 0-0 S(3) transition at 9.7 \um, which is located within the wide-band silicate absorption feature at the same wavelength. Consequently, A$_V$ can be estimated by examining the alignment of the 0-0 S(3) column density with respect to the column densities of the 0-0 S(2) and 0-0 S(4) lines in the assumption of OPR=3. 2) We considered line pairs originating from the same upper level, as their intrinsic ratio only depends on their frequency and radiative rates. Therefore, the observed ratio depends only on the extinction value at the considered wavelength ($A_{\lambda}$). This latter method can be applied only to regions covered by the NIRSpec FoV, as no transition from the same upper level falls in the MIRI range alone. 
A$_V$ was estimated by adopting the average of the values derived from each method, with typical uncertainties of the order of 1–2 mag. In cases where only MIRI data were available, the extinction was derived using solely the silicate absorption feature as an indicator. 
The extinction calculated using NIRSpec data is more robust, as it traces the excited gas, eliminating any dependence on the OPR. Moreover, it is less sensitive to the choice of the extinction law since, unlike the silicate absorption region, the extinction curve in the near-IR is better constrained.

We generated an extinction grid from A$_V=2$ to $30$ mag in steps of $0.1$, identifying the A$_V$ value that minimized the chi-squares ($\chi^2$) for the above mentioned methods, considering the corresponding errors. 
The extinction law of \cite{Pontoppidan24} has been adopted. For comparison, we also repeated the analysis using the extinction law of \citet{Gordon2023}.
The resulting parameters show only modest variations: on average, $T_{\rm warm}$ and $T_{\rm hot}$ differ by less than 10\%, while column densities ($N_{\rm H_2\text{-}warm}$, $N_{\rm H_2\text{-}hot}$) and $A_V$ vary by less than 20\%.
These differences do not affect the general trends or the physical interpretation discussed in the paper, confirming the robustness of our analysis with respect to the adopted extinction law.

\begin{figure}[h!]
    \centering
\includegraphics[width=0.45\textwidth,keepaspectratio]{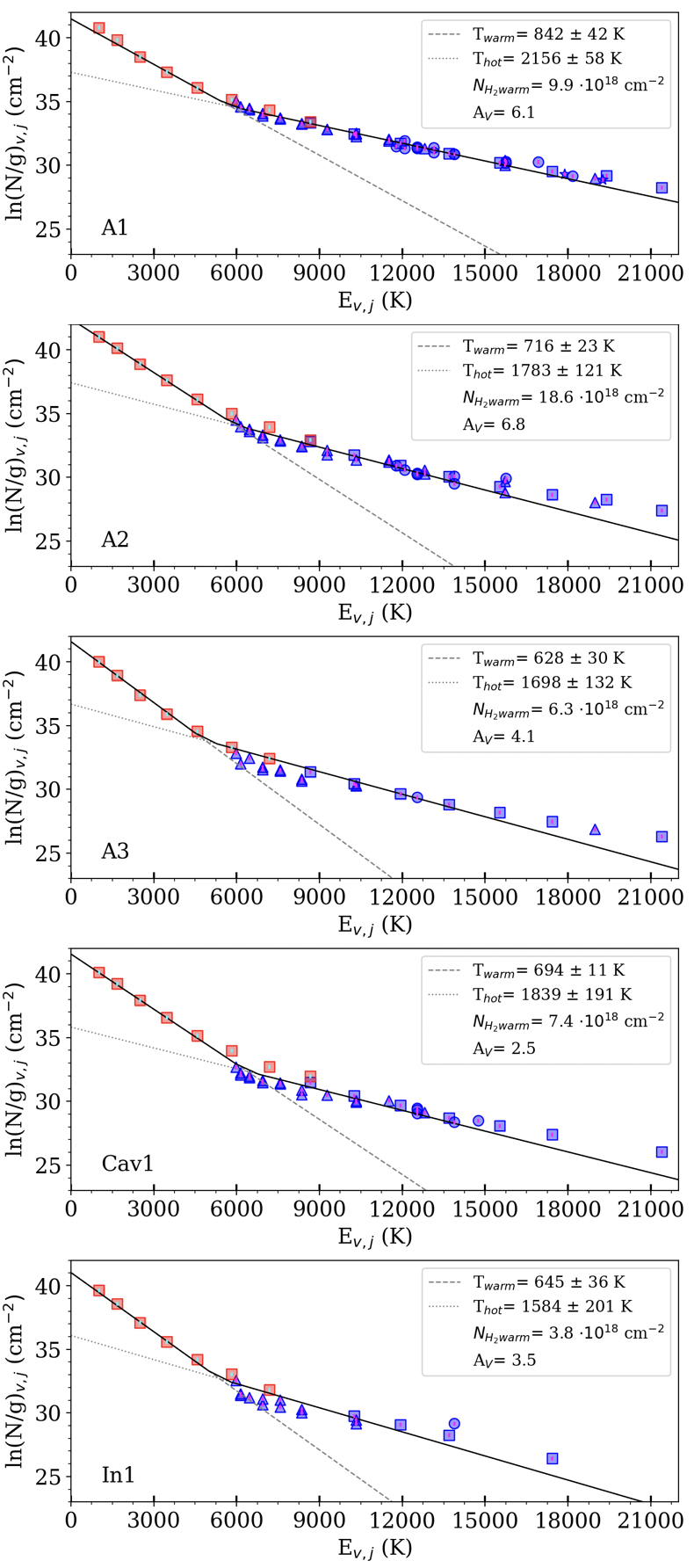}
\caption{Rotational diagrams with extinction-corrected column densities for regions in the blue-shifted outflow lobe covered by both MIRI and NIRSpec observations. 
Red and blue symbols correspond to MIRI and NIRSpec data, respectively. 
Symbols denote transitions from different vibrational levels: $v=0$ (squares), $v=1$ (triangles), $v=2$ (circles), and $v>2$ (stars).  
Straight lines indicate the two-component linear fits performed for each region. 
}
    \label{fig:rotationaln}
\end{figure}

The observed rotational diagrams deviate from a single straight line, showing a break around E$_{up}\sim$ 5000 K (Figure \ref{fig:rotationaln}). To account for this, we adopted a two-component model, i.e. we separately fitted 1) a warm component fitted to transitions with  E$_{up}< $5300 K, and 2) a hot component fitted to transitions with E$_{up}> $ 3700 K. The overlapping range between E$_{up}=$3700 - 5300 K is included in both fits to ensure continuity and robust fitting across the transition region.
In practice, the S(1)-S(5) lines trace the warm component, and the S(5)-S(8) and all NIRSpec lines trace the hot component. 


Figure \ref{fig:rotationaln} displays the rotational diagrams for the regions covered by both MIRI and NIRSpec observations represented by red and blue colors, respectively. 
Straight lines show the double linear fit applied to each region, with the corresponding parameters listed in Table \ref{table:bd}.
This table reports the column densities of the warm component of ($N_{\rm H_2-warm}$), which represents the bulk of the \htwo\ column density, and of the hot component ($N_{\rm H_2-hot}$), which is at least one order of magnitude lower.

Figure \ref{fig:rotationaln} confirms a posteriori the assumption of an OPR = 3 for these regions, as the pure rotational v=0 lines (squares) lack the zigzag pattern typically seen when the OPR ratio is much lower than the equilibrium value \citep[e.g.][]{neufeld2006}.

In A1 and A2, the observed column densities of the v=0 levels almost overlap  with the column densities of the higher vibrational levels with similar E$_{up}$. 
Considering the higher critical densities of the v$>$0 lines relative to the pure rotational transitions, and assuming collisional excitation as the dominant mechanism, this result suggests that the ro-vibrational levels have reached thermalization at densities $>$ 10$^{6}$ \cmt \citep{nisini2010}.

In the other considered regions, a slight misalignment is observed, more evident in the Cav1 and In1, suggesting an incomplete LTE for higher vibrational levels \citep[see Figure 7 of][]{nisini2010}. 
The detection of only a few lines with v$>$2, and their alignment with the transitions from lower vibrational levels, suggests that UV pumping does not play a dominant role in the observed \htwo\ emission. In fact, in a pure UV radiative excitation, high vibrational levels are populated by cascade from the excited electronic levels, and the corresponding rotational diagrams are characterized by a marked non-thermal distribution \citep{black1987,kaplan2021}.


The derived temperatures for both the warm and the hot components are highest at the emission peak A1, where $T_{\text{warm}} \sim$ 842 K and $T_{\text{hot}} \sim$ 2156 K. The cavity regions exhibit lower values of both temperature components ($T_{\text{warm}} \sim$ 600 - 700 K and $T_{\text{hot}} \sim$ 1500 - 1800 K).
Extinction values range from A$_V\sim2.5$ within the cavity—where lower extinction is expected—up to A$_V\sim$7 in the knots located near the central star, which is more embedded in the dusty envelope.

However, we observe that high-energy transitions (E$_{up} >$ 15000 K) are systematically underestimated by the fit obtained for the hot component, indicating the presence of an additional, higher-temperature component. Furthermore, the S(1) line of the A1 knot deviates above the warm component fit, indicating the presence of a lower-temperature component.
This evidence suggests a multi-temperature distribution, reflecting a more complex temperature stratification in the emitting gas.

Table \ref{table:bd} also presents the results obtained in regions where only MIRI data are available, 
marked with the $\dag$ symbol. 
The derived values of the warm component are consistent with the procedure applied in regions where NIRSpec data are included. However, the hot component is underestimated because of the absence of lines excited at higher energy levels. The A$_V$ values were derived relying exclusively on the \htwo\ S(3) transition at 9.7 \um\ and are systematically higher than those derived in the other regions, especially in the red-shifted lobe where values up to $\sim$30 mag are reached in the inner regions.  
 
\begin{table*}[t]
\caption{Parameters obtained from the fit of rotational diagrams. The first five columns report the double-linear fit parameters $T_{\text{warm}}$, $T_{\text{hot}}$, $N_{\rm H_2-warm}$, $N_{\rm H_2-hot}$ and A$_V$ (see Figure \ref{fig:rotationaln}). The last five columns report the parameters obtained using the temperature stratification approach ( $T_{\text{min}}$, $N_{\rm H_2}$, A$_V$, $\beta$ and OPR, see Figure \ref{fig:temp_stra} and \ref{fig:ap_temp_str_MIRIALL}). }
\label{table:bd}
\begin{tabular}{c|c c c c c |c c c c c}
\hline \hline
Model & \multicolumn{5}{c|}{Two component fit}  & \multicolumn{5}{c}{Temperature stratification} \\ \hline
\hline
Region & $T_{\text{warm}}$ & $T_{\text{hot}}$ & $N_{\rm H_2-warm}$ & $N_{\rm H_2-hot}$ & A$_V$ & $T_{\text{min}}$ & $N_{\rm H_2}$ & A$_V$ & $\beta$ & OPR \\
 &{\small ($K$)} &{\small ($K$)}  &  {\small ($10^{18}\text{cm}^{-2} $)}  &  {\small ($10^{17}\text{cm}^{-2} $)} & & {\small ($K$)} & {\small ($10^{18}\text{cm}^{-2} $)}  \\ 
\hline
A1 & 842 ± 42 & 2156 ± 58 & 9.9 & 2.7 & 6.1 &  400  &  50.1  &  6.5  &  4.0  &  3.0 \\ 
A2 & 716 ± 23 & 1783 ± 121 & 18.6 & 1.5 & 6.8 &  500  &  50.1  &  5.9  &  5.2  &  3.0 \\ 
A3 & 628 ± 30 & 1698 ± 132 & 6.3 & 0.5 & 4.1 &    350  &  20.0  &  0.7  &  4.6  &  3.0 \\ 
Cav1 & 694 ± 11 & 1839 ± 191 & 7.4 & 0.3 & 2.5 &  500  &  20.0  &  0.5  &  5.6  &  3.0 \\ 
In1 & 645 ± 36 & 1584 ± 201 & 3.8 & 0.2 & 3.5 &  450  &  12.6  &  2.7  &  5.4  &  3.0 \\

A4$^\dag $ & 661 ± 12 & 1253 ± 28 & 12.1 & 1.7 & 5.0 &  450  &  31.6  &  2.7  &  4.8  &  3.0 \\ 
A5$^\dag $ & 618 ± 23 & 1406 ± 7 & 13.4 & 1.4 & 6.7 &  350  &  31.6  &  0.9  &  4.2  &  2.7 \\ 
A6$^\dag $ & 626 ± 17 & 1221 ± 16 & 11.6 & 1.1 & 5.2 &  400  &  31.6  &  1.9  &  4.8  &  2.7 \\ 
Cav2$^\dag $ & 809 ± 26 & 1268 ± 60 & 28.7 & 2.4 & 26.9 & 500  &  50.1  &  20.9  &  5.8  &  2.7 \\ 
Cav3$^\dag $ & 659 ± 10 & 1585 ± 1585 & 13.4 & 0.7 & 19.7 &  150  &  79.4  &  6.1  &  3.6  &  1.5 \\ 
Bow1$^\dag $ & 600 ± 15 & 1435 ± 16 & 18.4 & 1.0 & 13.3 & 300  &  50.1  &  5.5  &  4.4  &  2.4 \\ 
Bow2$^\dag $ & 547 ± 8 & 1221 ± 1221 & 27.2 & 0.6 & 15.9 &  200  &  125.9  &  3.1  &  4.4  &  1.5 \\ 
In2$^\dag $ & 678 ± 23 & 1348 ± 1348 & 16.1 & 0.6 & 24.2 &  100  &  199.5  &  7.9  &  3.6  &  1.2 \\ 
 \hline
\end{tabular}
\centering
\\ \small  $\dag $ Regions where the rotation diagrams contain only MIRI data.\\
\end{table*}

\subsubsection{Temperature stratification} \label{sec:tempstr}
Motivated by the detection of a clear temperature stratification in the rotational diagrams, potentially originating in post-shock regions, we applied a temperature gradient approach. 
Specifically, we adopted the approach presented in \cite{neufeld&yuan2008} and applied to \spitzer\ data by \cite{neufeld2009,nisini2010} and \cite{giannini2011}, assuming a slab of gas where the column density in each layer at a given $T$ varies as $ dN \propto T_{\text{ex}}^{-\beta} $. Optically thin LTE is also assumed in each of the considered layers. 
This law is integrated, to find the total \htwo\ column density $N_{\rm H_2}$, between a $T_{\text{min}}$ and a $T_{\text{max}}$. We have kept $T_{\text{max}}$ fixed to 4000 K as higher temperature gas is not expected to contribute to the line emission in our observed wavelength range. Therefore, the parameters of the model are $T_{\text{min}}$ (which we varied between 100 and 400 K), the exponent $ \beta $, $N_{\rm H_2}$ and the visual extinction A$_V$ by which the modeled column densities have been reddened before comparing them to the observed values. We also consider the OPR as a free parameter to explore whether deviations from the equilibrium value are observed in any of the considered outflow positions. 


The best fit for the 13 regions, listed in Table \ref{table:bd}, has been obtained through a $\chi^2$ minimization procedure. 
We applied our model exclusively to the v=0-0 transitions because, as discussed in Subsection \ref{sec:dfit}, transitions from higher vibrational 
levels could be sub-thermally excited.
This means that we consider transitions from S(1) to S(15), while we only fit transitions from S(1) to S(8) in regions observed by MIRI only.

\begin{figure}[t]
    \centering
\includegraphics[width=0.5\textwidth,keepaspectratio]{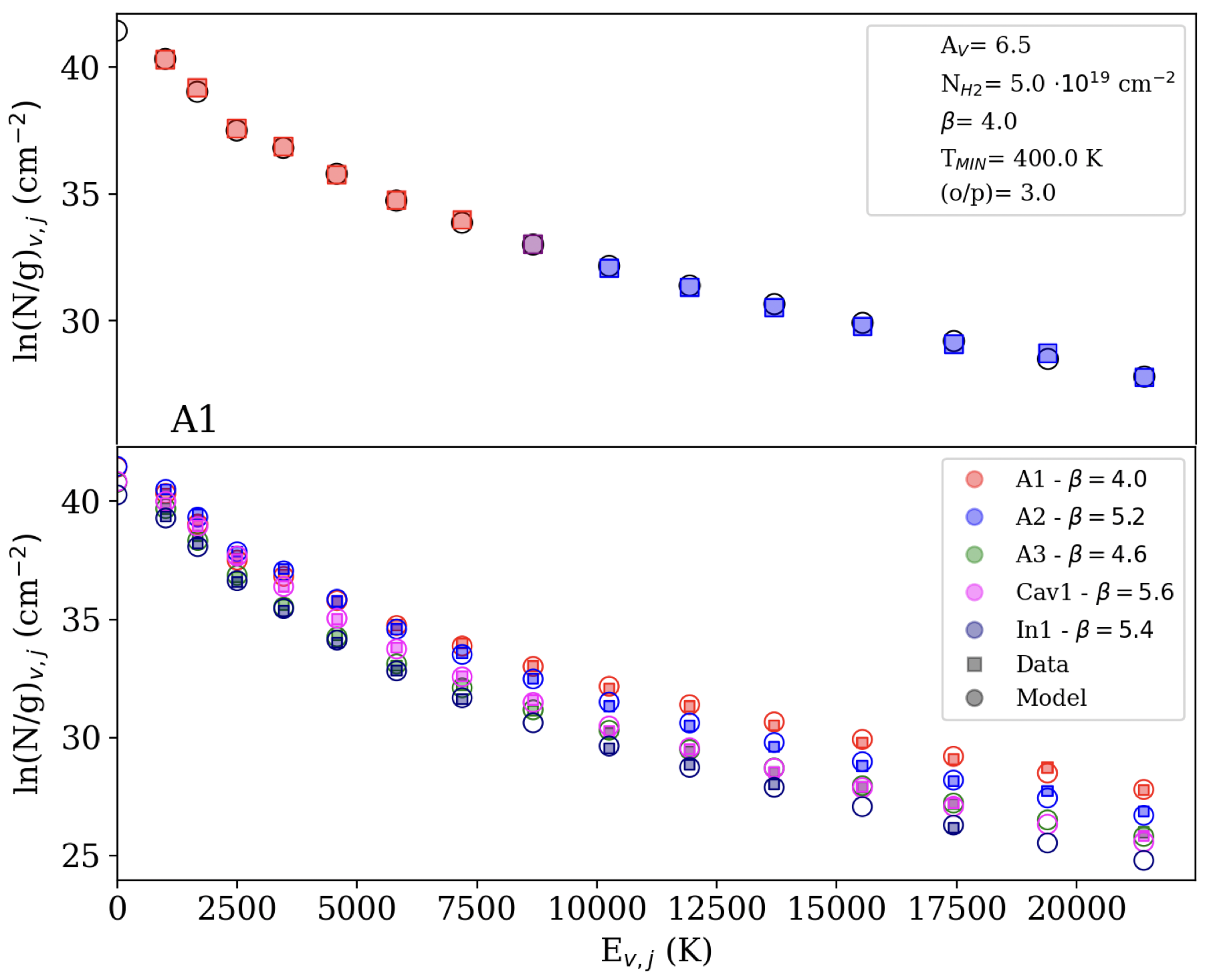}
\caption{
\textit{Upper panel:} Rotational diagram of the pure rotational $v=0$ lines for knot~A1, assuming a temperature stratification. 
Red and blue squares represent MIRI and NIRSpec data, respectively, while black open circles denote the best-fit model. 
The corresponding fitted parameters are listed in the inset. 
\textit{Lower panel:} Rotational diagrams of the pure rotational $v=0$ lines for all regions covered by both NIRSpec and MIRI. 
Colored squares represent the observed data for each region, and open circles show the corresponding best-fit models.}
    \label{fig:temp_stra}
\end{figure}

The upper panel of Figure \ref{fig:temp_stra} presents the result for knot A1. 
Plots of the remaining regions covered by NIRSpec observations are shown in Figure \ref{fig:temp_str_reg} in Appendix and the derived parameters for each region are detailed in Table \ref{table:bd}. The lower panel of Figure \ref{fig:temp_stra} shows the results for all five regions covered by MIRI and NIRSpec, with different colors representing each region.
$ \beta $ range between 4 and 5.6. 
Lower values of $\beta$ indicate a relatively larger fraction of gas at higher temperatures. Consistently, we find the lowest $\beta$ in knot A1, which also shows the highest hot temperature component based on the linear fit.
$T_{\text{min}}$ is lower than the temperature obtained for the warm component in the double linear fit ($T_{\text{warm}}$) which is expected since the temperature stratification fit takes into account also colder gas not traced by the S(1)-S(5) transitions. 
The OPR is consistent with the equilibrium value of 3, as observed in the double linear fit. The total column density is similar for the five regions, ranging from $N_{\rm H_2}\sim 1.2-5.0 \times 10^{19}$ \cmd. These values are nearly an order of magnitude higher than those derived from the linear fit (Figure \ref{fig:rotationaln}, Table \ref{table:bd}), which is expected since the model includes the contribution from the coldest gas not traced by MIRI.



\begin{figure*}[ht]
    \centering    \includegraphics[width=1\textwidth,keepaspectratio]{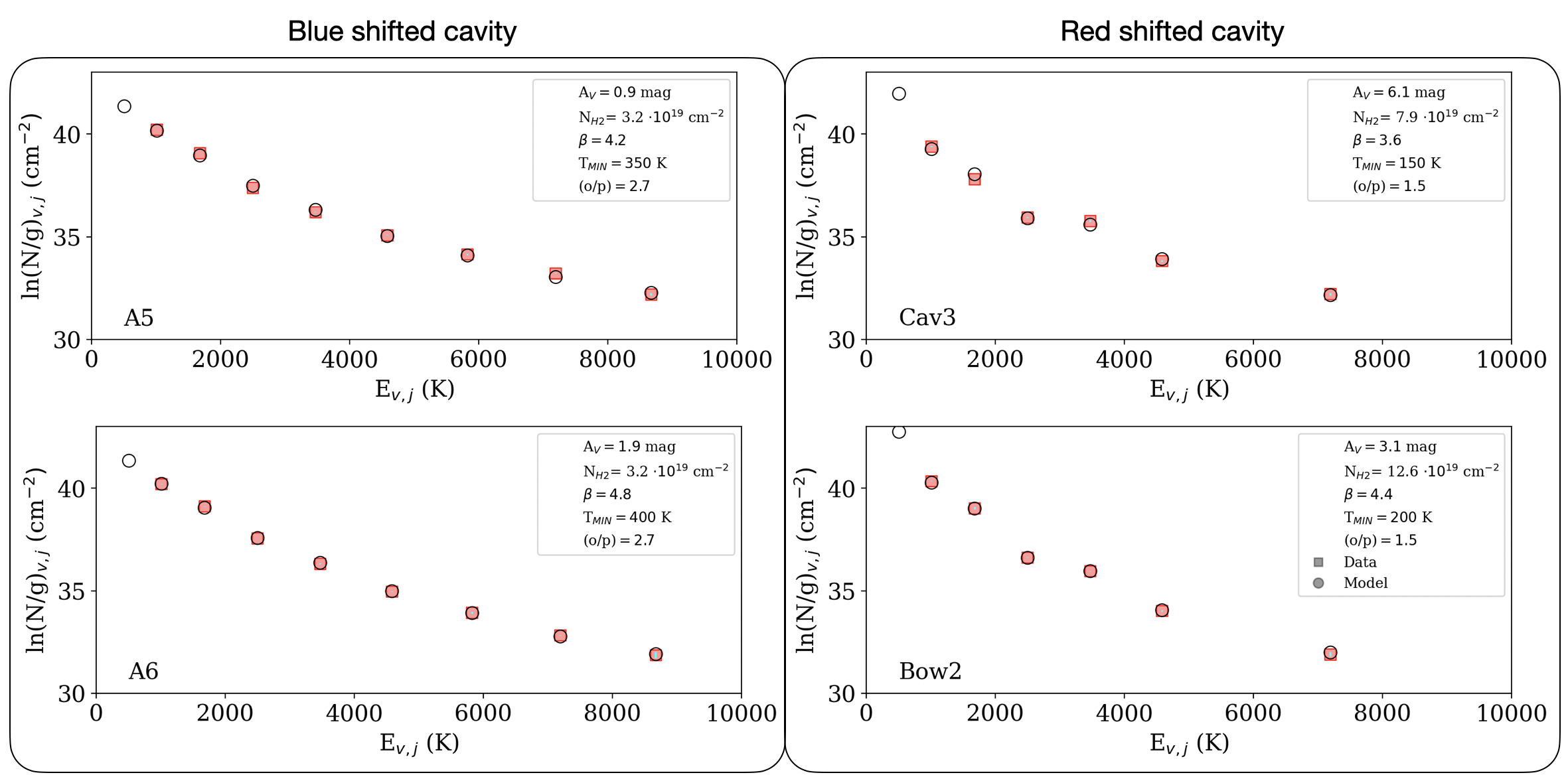}
    \caption{
Rotational diagrams of the pure rotational $v=0$ H$_2$ lines observed with MIRI in representative regions of the red-shifted (A5 and A6) and blue-shifted (Cav3 and Bow2) outflows. 
Filled squares represent the MIRI data, while black open circles denote the best-fit models.}
    \label{fig:temp_stra_miri}
\end{figure*}

A similar analysis was conducted for the outer blue-shifted and red-shifted regions, not covered by the NIRSpec observations, employing the MIRI-covered lines only. 
Figure \ref{fig:temp_stra_miri} presents the results for four representative regions in the blue- and red-shifted outflow regions, with additional regions shown in Figure \ref{fig:ap_temp_str_MIRIALL} in Appendix. 
In the blue-shifted region, the fitted parameters are consistent with those derived from the areas covered by NIRSpec. Conversely, the red-shifted regions exhibit higher column densities, increased extinction, and OPR that deviate from the equilibrium value in some locations. Notably, regions with the lowest OPR values (OPR=1.2) also correspond to the lowest minimum temperatures $T_{\text{min}}=100$ K), suggesting that the gas may not have had sufficient time to reach OPR equilibrium (see discussion in Section \ref{sec:disc}).


When comparing results obtained with different methods, we note that A$_V$ calculated applying the temperature stratification approach is in most cases lower than those derived in the double linear fit. This is because the curvature introduced by the stratification predicts a lower column density on the S(3) line compared to the linear model used in the double fit. This trend is reflected in Table \ref{table:bd},  with a few exceptions. 
Notably, the very high extinction values found in regions like Cav2 cast doubt on the reliability of the fit. Another exception is observed in the red-shifted outflow regions with low OPR, such as Cav3 and In2, where the linear fit does not account for the characteristic zigzag pattern associated with these OPR values, leading to less accurate extinction estimates.

The differences in the column density values obtained in the red-shifted regions are more pronounced compared to those located in the blue-shifted part. The higher $N_{\rm H_2}$ derived from the temperature stratification method in the red-shifted region results from the higher contribution from colder gas with $T<T_{\text{warm}}$.


\subsection{Maps of parameters}

Motivated by the complex internal structure of the outflow, we applied the analysis of the rotational diagrams pixel by pixel, using line intensities derived from line maps to build spatially resolved maps of temperature ($T$), column density ($N_{\rm H_2}$), and extinction (A$_V$). These maps were constructed exclusively from MIRI data due to their broader FoV, which captures the various outflow structures. Therefore, the analysis is limited to the purely rotational 0-0 S(1)-S(8) lines, and consequently does not probe the components at the highest temperature (blue squares in Figure \ref{fig:rotationaln}).

Since the MIRI spatial resolution and FoV vary across sub-bands and wavelengths, all line maps were regridded to match the lowest-resolution sub-band (Channel 3C) at 17 \um, corresponding to a FWHM of $\sim$0.\asec6, thereby ensuring a consistent analysis across all transitions.
The chosen reprojection algorithm, $reproject\_adaptive$, is part of the Astropy-affiliated reproject package \citep{reproject}, which uses an adaptive anti-aliased resampling technique described in \citet{DeForest04}.  We enabled the $conserve\_flux$ flag to preserve total flux across resampling.

As can be seen from the rotational diagrams of Figures \ref{fig:rotationaln}, \ref{fig:temp_stra} and \ref{fig:temp_stra_miri}, the column densities of the MIRI lines do not conform to a single straight line but already show the curvature relative to the temperature stratification. 
However, due to the low SNR for the lines of each pixel, especially the high-J lines in the red-shifted lobe, the application of the temperature stratification model is challenging, resulting in a high degeneracy of the parameters.   
For this reason, we applied the double-temperature linear fit approach, separately fitting the low- and high-temperature components as described in Subsection~\ref{sec:dfit}. Since NIRSpec lines are not available, the extinction was derived solely from the flux attenuation of the S(3) line within the silicate absorption feature, following the method described in Subsection~\ref{sec:dfit}. In this case, the energy cutoff between the two components was set at $E_{\rm up} = 4000$~K, slightly lower than that adopted for the NIRSpec analysis. This choice accounts for the smaller number of detected transitions (eight lines in total) and ensures that both temperature components are adequately sampled. The analysis was restricted to lines with SNR$>$5 and to pixels with at least three detected lines in each temperature regime, ensuring the robustness of the linear fit. In practice, the S(1)–S(4) lines trace the warm component, while the S(5)–S(8) lines trace the hot component.
The analysis was restricted to lines with SNR$>$5 and to pixels with at least three detected lines in each temperature regime, ensuring the robustness of the linear fit.

\begin{figure*}[ht]
    \centering
\includegraphics[width=0.9\textwidth,keepaspectratio]{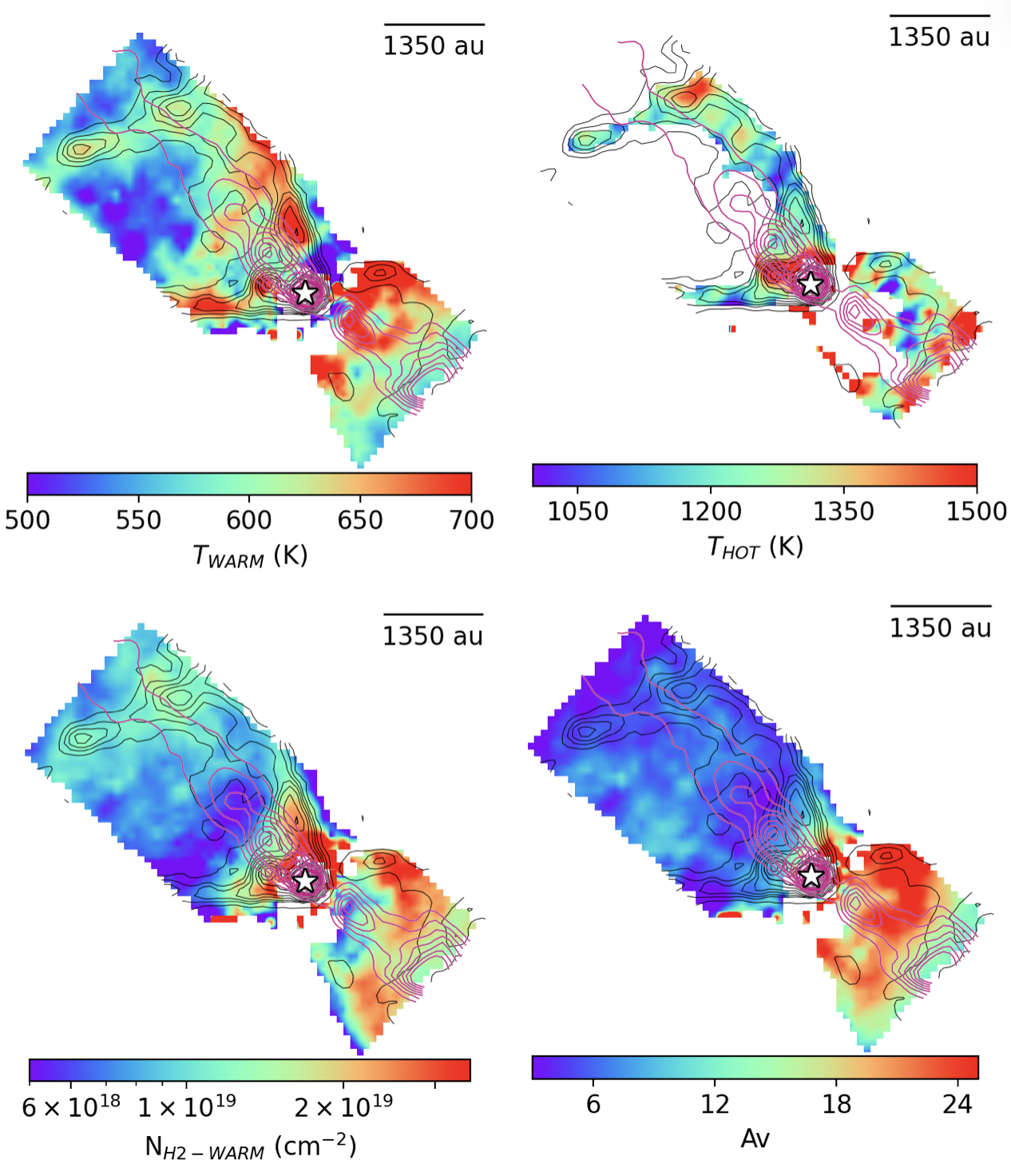}
\caption{
(\textit{Top left}) Temperature map of the warm H$_2$ component, derived from the 0--0~S(1)–S(4) transitions. 
(\textit{Top right}) Temperature map of the hot H$_2$ component, obtained from the 0--0~S(5)–S(8) transitions. 
(\textit{Bottom left}) H$_2$ column density map of the warm component. 
(\textit{Bottom right}) Visual extinction (A$_V$) map. 
White stars indicate the source position. 
Magenta contours correspond to the \feii\ 5.34 \um\ line tracing the atomic jet (levels from 0.4 to 17.0~MJy~sr$^{-1}$), 
while black contours represent the H$_2$ 0--0~S(7) 5.51~\um\ line (levels from 0.4 to 4.0~MJy~sr$^{-1}$). 
Only lines with SNR~$>$~5 were included in the analysis; pixels with fewer than three detected lines per temperature regime were excluded. 
The maps were convolved to match the spatial resolution of MIRI–MRS at 17~\um, corresponding to a FWHM of $\sim$0\farcs6 and a spectral resolving power of $R \sim 1600$–$1800$ ($\Delta v \approx 170$ \kms; \citealt{Argyriou2023}; \citealt{Wells15}; JWST Documentation). 
This wavelength provides the lowest spatial and spectral resolution among the H$_2$ lines used to construct the rotational diagrams.}
    \label{fig:allmaps}
\end{figure*}

The upper panels of Figure \ref{fig:allmaps} display the temperature maps for the warm ($T_{\text{warm}}$) and hot ($T_{\text{hot}}$) components. 
The $T_{\text{warm}}$ map shows slightly higher temperatures near emission knots and around the source, with minimal variation elsewhere, ranging between $500$ and $700$ K, thus $\Delta T \sim 200$ K.
The $T_{\text{hot}}$ map instead reveals more pronounced temperature changes, with $\Delta T \sim 500$ K. The A1 knot appears more excited than its surroundings, and a high-temperature region is observed at the edge of the red-shifted outflow, 
corresponding to the large bow-shock created by the atomic jet. 
Both maps detect higher temperatures also at the cavity walls and lower temperatures inside the cavity, following the line brightness distribution without strong gradients.

The $N_{\rm H_2-warm}$ column density map is shown in the lower-left panel of Figure \ref{fig:allmaps}, illustrating the distribution of \htwo\ molecules along the line of sight. 
Column densities peak at the cavity’s base, where gas thickness is maximized due to geometrical effects and compression by shocks. In the blue-shifted outflow the column density is higher along the cavity walls, with higher values in the northern region.
In the red-shifted outflow, the highest column density appears at the cavity edge near the bow shock, suggesting material laterally pushed and compressed by the propagation of the atomic jet.

Lower-right panel of Figure \ref{fig:allmaps} presents the A$_V$ map. Extinction values range between a few mag in the interior of the NE cavity and up to about 25 mag at the edge of the SW cavity. The red-shifted outflow lobe naturally shows the highest extinction, which decreases around the bow shock as the atomic jet has swept up the intervening material. However a word of caution need to be considered for the derived A$_V$ in some of the red-shifted regions since the assumption of OPR=3 might be not always valid, as we have show in Section \ref{sec:tempstr}.
In the blue-shifted lobe, we observe a lane of higher extinction that is not aligned with any outflow structures, likely due to dust inhomogeneity in the foreground cloud. 


\subsection{Kinematics}

We investigated the velocity structure of the outflow by performing Gaussian fits to the line profiles of the brightest transitions at each spatial position. Although the spectral resolution of our observations is limited, the radial velocity of the brightest lines can be determined with higher precision, which scales with the line SNR as $\Delta v_{\text{inst}}/$SNR, where $\Delta v_{\text{inst}}$ is the instrumental resolution \citep{porter2004}. 
Radial velocity maps were then produced from the continuum-subtracted data cubes by fitting the selected lines on a pixel-by-pixel basis, excluding spectra with line fluxes below SNR$<$10.
We generated radial velocity maps using the continuum-subtracted sub-cubes and performing a Gaussian fit of the lines on a pixel-by-pixel basis. Line fluxes with SNR$<$10 were ignored. 

Radial velocities were first corrected from the Solar System barycentric frame to the Local Standard of Rest (LSR) using the IRAF routine $rvcorrect$, and subsequently shifted by +5 km \kms\ to account for the systemic velocity of the parent cloud \citep{arce2013}. Velocity maps were then constructed for the 1–0 S(1) line at 2.12 \um\ in the NIRSpec range and for the 0–0 S(1)–S(8) lines in the MIRI range.

The error associated with the velocity maps combines Gaussian‐fit uncertainties and the absolute wavelength calibration accuracy, which dominates and varies by instrument: $\sim15$ \kms\ for NIRSpec high‑resolution, and a few \kms\ in MIRI‑MRS bands 1A–3B rising to $\sim 30$ \kms\ in bands 3C–4C \citep{stsci_miri_mrs_calibration}.

\begin{figure*}[t]
    \centering
\includegraphics[width=1\textwidth,keepaspectratio]{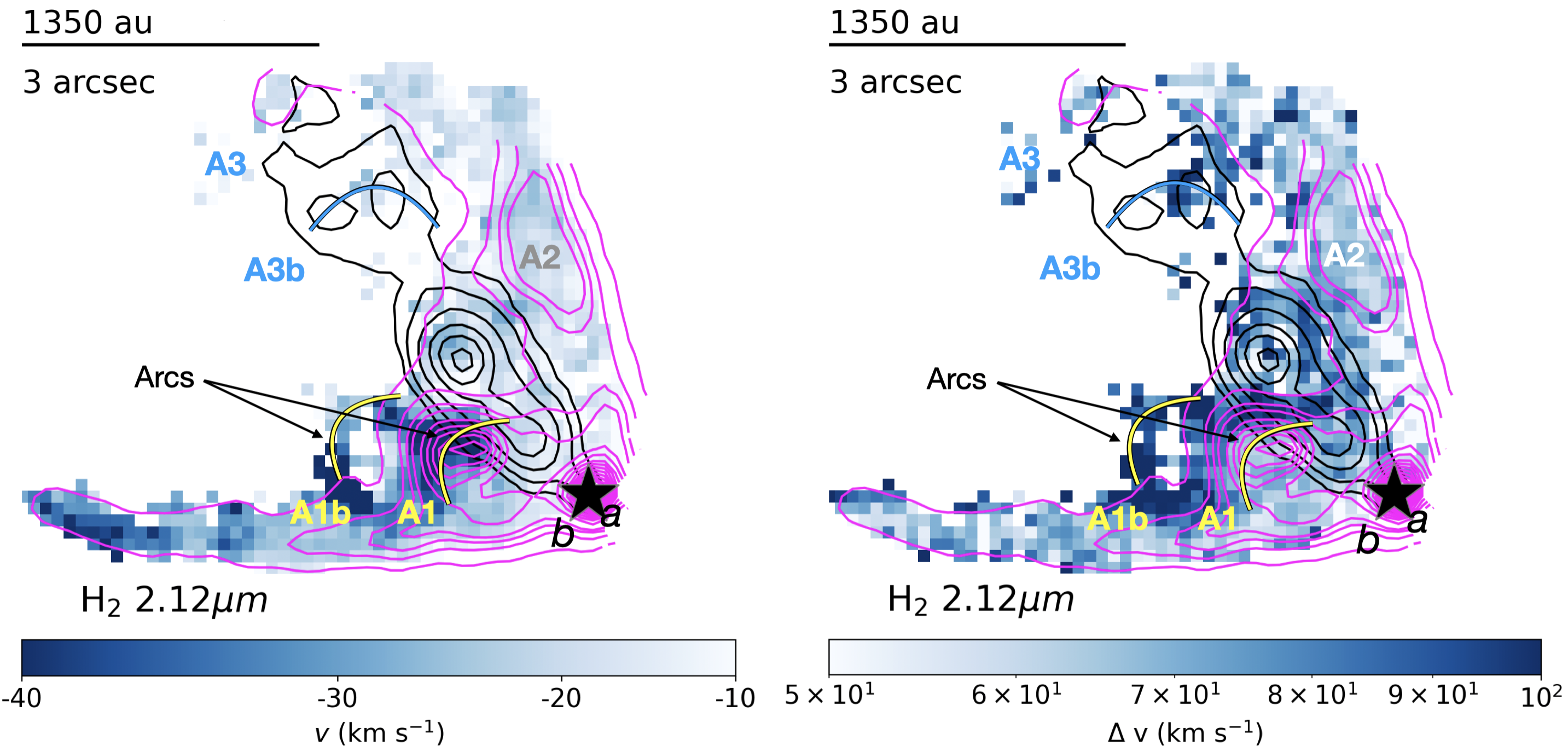}
\caption{
(\textit{Left}) Radial velocity map of the H$_2$ 1--0~S(1) line at 2.12~\um\ obtained with NIRSpec. 
(\textit{Right}) FWHM map in velocity space derived from the same line. 
Velocities are given in the rest frame of the ambient CO cloud. 
Black stars indicate the positions of the binary sources. 
Magenta contours trace the stacked H$_2$ emission (levels from 0.085 to 2.0~MJy~sr$^{-1}$), while black contours correspond to the [Fe\,\textsc{ii}] 1.81~\um\ line delineating the collimated atomic jet (levels from 0.028 to 0.4~MJy~sr$^{-1}$).}
    \label{fig:vel_nirspec}
\end{figure*}

Left panel of Figure \ref{fig:vel_nirspec} presents the radial velocity map of the 1–0 S(1) 2.12 \um\ line within the blue-shifted region of the inner outflow. Radial velocities range from $v= -10$ \kms\ to $v= -40$ \kms\, consistent with earlier studies by \cite{birney2024}, using SINFONI data with higher spectral resolution (R=4000).
In particular, we confirm the northern side of the flow cavity displaying a lower velocity ($|v|\sim$10-20 \kms) with respect to the southern part where velocities reach $v\sim$30-40 \kms\ in absolute value. 

\cite{birney2024} discussed the origin of this velocity asymmetry. 
Flow rotation was excluded on the basis of the unreliably large specific angular momentum implied by the observed \htwo\ velocity gradient. 
Furthermore, the direction of the apparent rotation is opposite to that of the CO envelope emission, further arguing against a rotational origin.
Instead, \cite{birney2024} suggest that it could be due to the molecular outflow expanding in an inhomogeneous ambient medium or the presence of two separate outflows from the binary system. 
Our NIRSpec observations at higher spatial resolution are in favor of this latter hypothesis. Indeed, we reveal substructures in the velocity map that remained undetected at the seeing limited SINFONI observations. In particular, the velocity in knot A1 appears resolved into two bow-shocks at velocities $v\sim$-40 \kms\, corresponding to the two \htwo\ emission shells (A1 and A1b) highlighted in Figures \ref{fig:nirspec_linemap} and \ref{fig:vel_nirspec}, and pointing towards the source b$^\star$. 
Interestingly, if arc A1b is interpreted as a bow shock, the velocity appears higher along its wings than at the apex, contrary to expectations. In jet-driven bow shocks, the maximum velocity is typically reached at the bow apex, while it progressively decreases toward the flanks.
A gradient in velocity along the southern cavity is also observed, ranging from $v\sim -20$ to $v\sim -40$ \kms.
The latter could also be due to the expanding \htwo\ wind interacting with the cavity wall, or, alternatively, it could trace the broad wings of previous, older bow-shock piling-up against the cavity walls \citep[e.g.][]{Rabenanahary2022}. 

Right panel of Figure \ref{fig:vel_nirspec} shows the map of the intrinsic Full Width at Half Maximum (FWHM) in velocity space, obtained by subtracting in quadrature the instrumental width.
The typical uncertainty in the fitted FWHM for NIRSpec spectra with SNR$>10$ is generally estimated to range between 5 and 15 \kms, depending on the instrument’s spectral resolution and the sampling of the data \citep{NIRSpecHandbook2024}. Given that the pixels in our map have SNR$>>10$, we find uncertainties in the FWHM well below 10 \kms.
The intrinsic FWHM map presents values that vary along the outflow, overall exceeding $\Delta v>50$ \kms, indicating that the lines are resolved. These high FWHM values in the highlighted region consistent with shock excitation at the origin of the observed \htwo\ emission.

\begin{figure}[t]
    \centering
\includegraphics[width=0.5\textwidth,keepaspectratio]{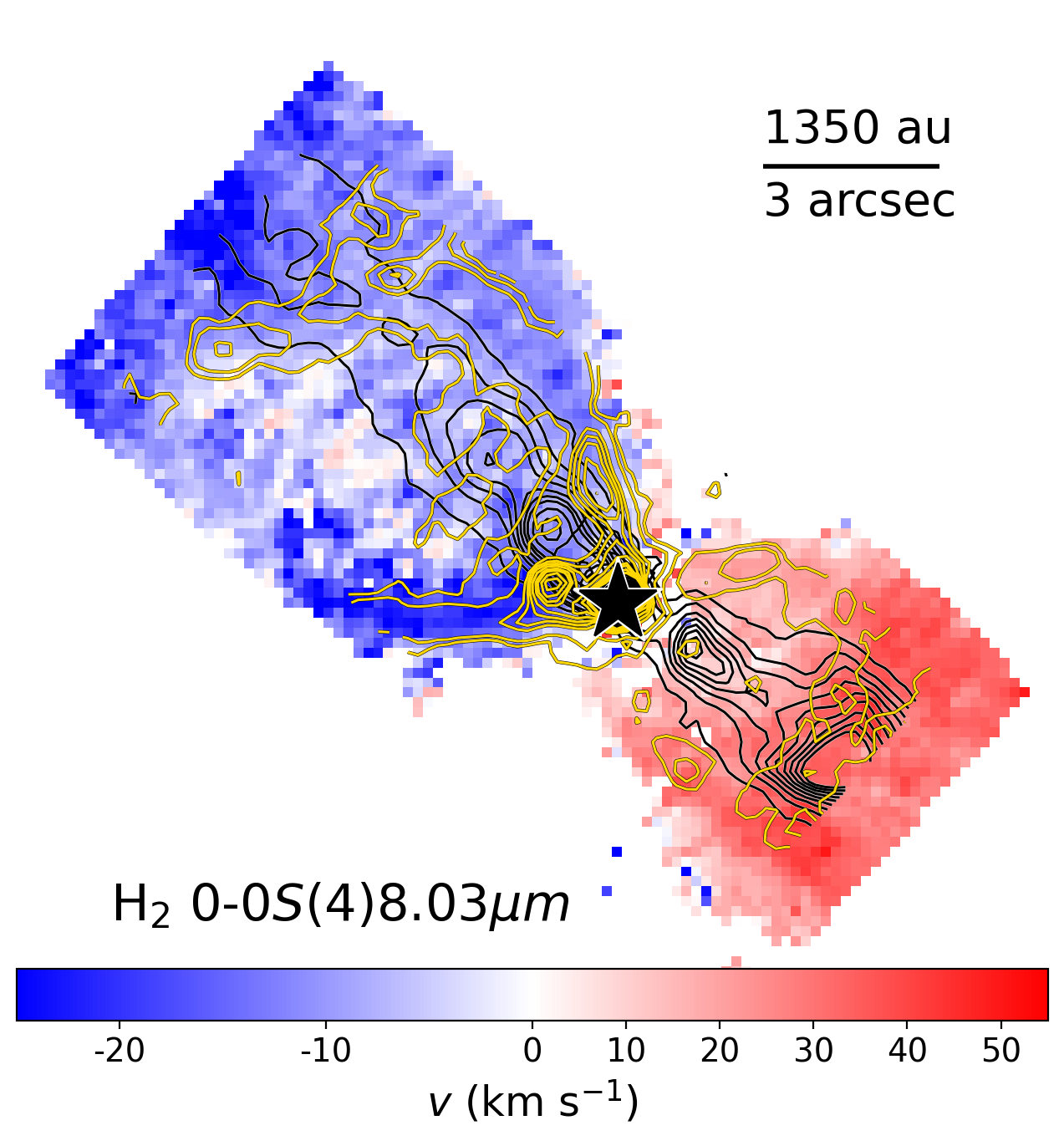}
    \caption{\htwo\ radial velocity map obtained with MIRI 0-0 S(4) line at 8.08 \um. Velocities are in the rest frame of the ambient CO cloud. Black stars mark the source position.
    Yellow and black contours correspond to the \htwo\ emission traced by the S(4) line at 8.03 \um\ (levels from 0.4 to 4.0 MJy sr$^{-1}$) and collimated atomic jet emission traced by the \feii\ at 1.81 \um\ (levels from 0.4 to 17.0 MJy sr$^{-1}$), respectively.
    }
    \label{fig:vel_miri8}
\end{figure}

Figure~\ref{fig:vel_miri8} shows the MIRI velocity map of the H$_2$~S(4) line at 8.08 \um. This transition was selected as an optimal compromise between spectral resolving power (R $\sim 100$ \kms) and SNR ratio, since lines at shorter wavelengths are too heavily affected by extinction in the redshifted lobe. Moreover, the spectral resolution at 8.08 \um\ is comparable to that of the NIRSpec data, enabling a consistent comparison between the two instruments.


The morphology of the 8.08 \um\ velocity distribution is consistent with that of the 2.12 \um\ in the NIRSpec overlapping blue-shifted area: higher velocities in the southern arm of the cavity compared to the northern one. We however note that the absolute value of the radial velocities is different in the two lines: while the 2.12 \um\ line shows a maximum blue-shifted velocity of $v \sim$-40 \kms\ towards knot A1, the 8.08 \um\ velocity in the same position is $v\sim$-20 \kms. This difference is larger than the wavelength calibration errors and will be further addressed later in this section.

The velocity within the blue-shifted cavity appears fairly uniform and only slightly blue-shifted ($\sim -10$ \kms), whereas higher velocities are observed along the cavity edges. This pattern is likely a geometrical effect. The system consists of a wide-angle cavity inclined by $\sim$37 degree with respect to the line of sight, where the \htwo\ emission originates mainly from the interaction with the cavity walls. In such a configuration, the projected radial velocities within the cavity tend to approach zero, as the line-of-sight components from the approaching and receding walls largely cancel each other out.

The radial velocities map is not symmetric between the red-shifted the blue-shifted regions. We note, in particular, that the red-shifted velocity increases with the distance from the central source, which is a trend not observed in the blue-shifted portion. This is likely due to the interaction of the atomic jet with the molecular material. The high velocity jet pushes the dense ambient medium creating the large molecular bow-shock accelerated at radial velocity up to $v=40$ \kms\ . The radial velocity then decreases in the bow-wings closer to the source. We also observe a trend similar to the blue-shifted outflow, namely low velocity ($\sim$10-20 \kms) in the inner region, and higher velocity ($\sim$20-30 \kms) at the flow border.

Figure~\ref{fig:vel_maps_miri} in the Appendix presents the corresponding MIRI velocity maps for lines with different excitation energies. The overall morphology is consistent across all transitions, although the absolute velocity values vary slightly between maps. The color scale in Figure~\ref{fig:vel_maps_miri} is global—i.e., the same for all transitions—so any minor differences between the S(4) map shown in Figure~\ref{fig:vel_miri8} and that in Figure~\ref{fig:vel_maps_miri} arise solely from this difference in scaling.

\begin{table*}[t]
 \caption{Peak velocity and width of representative transitions at different excitation energies for the different outflow regions. } 
 \label{tab:kinematics} 
  \begin{tabular}{ c c c c c c c c c c } 
 \hline\hline 
  \multicolumn{1}{c|}{ }   & \multicolumn{2}{c|}{$2.12 \mu m$ }  & \multicolumn{2}{c|}{$5.51 \mu m$ }  & \multicolumn{2}{c|}{$8.03 \mu m$ }  & \multicolumn{2}{c}{$12.28 \mu m$ }   \\  
 \multicolumn{1}{c|}{Reg }   &  \multicolumn{1}{c}{$v_{peak}$ } &  \multicolumn{1}{c|}{$\Delta v$ }  &  \multicolumn{1}{c}{$v_{peak}$ } &  \multicolumn{1}{c|}{$\Delta v$ }  &  \multicolumn{1}{c}{$v_{peak}$ } &  \multicolumn{1}{c|}{$\Delta v$ }  &  \multicolumn{1}{c}{$v_{peak}$ } &  \multicolumn{1}{c}{$\Delta v$ }   \\  
 \multicolumn{1}{c|}{ }   & \multicolumn{2}{c|}{\small (km $s^{-1}$) }  & \multicolumn{2}{c|}{\small (km $s^{-1}$) }  & \multicolumn{2}{c|}{\small (km $s^{-1}$) }  & \multicolumn{2}{c}{\small (km $s^{-1}$)}   \\  
 \hline 
A1 & -39.5 $\pm$ 0.4 & 69.0 $\pm$ 0.9 & -24.4 $\pm$ 0.4 & 41.5 $\pm$ 0.8 & -19.6 $\pm$ 0.4 & 29.1 $\pm$ 1.0 & -14.1 $\pm$ 0.9 & 36.0 $\pm$ 1.9 \\ 
A2 & -19.7 $\pm$ 0.4 & 63.2 $\pm$ 0.9 & -10.7 $\pm$ 0.3 & 18.9 $\pm$ 0.7 & -8.1 $\pm$ 0.5 & 8.2 $\pm$ 1.1 & -9.9 $\pm$ 0.8 & 48.0 $\pm$ 1.9 \\ 
A3 & -14.3 $\pm$ 0.5 & 78.1 $\pm$ 1.9 & -13.2 $\pm$ 0.4 & 31.4 $\pm$ 1.1 & -8.0 $\pm$ 0.6 & 21.2 $\pm$ 1.5 & -8.1 $\pm$ 0.8 & 44.5 $\pm$ 2.0 \\ 
Cav1 & -27.2 $\pm$ 0.5 & 69.2 $\pm$ 1.0 & -22.2 $\pm$ 0.7 & 30.8 $\pm$ 1.5 & -20.1 $\pm$ 0.5 & 30.0 $\pm$ 1.0 & -19.7 $\pm$ 0.9 & 19.3 $\pm$ 2.2 \\ 
In1 & -18.6 $\pm$ 0.5 & 80.0 $\pm$ 1.2 & -14.4 $\pm$ 0.4 & 26.0 $\pm$ 0.8 & -6.6 $\pm$ 0.6 & 15.5 $\pm$ 1.3 & -12.2 $\pm$ 0.9 & 16.7 $\pm$ 2.1 \\ 
A4 $^\dag $ &  &   & -10.4 $\pm$ 0.4 & 29.4 $\pm$ 0.8 & -11.0 $\pm$ 0.5 & 11.8 $\pm$ 1.1 & -11.7 $\pm$ 0.8 & 55.0 $\pm$ 2.0 \\ 
A5 $^\dag $ &  &   & -14.7 $\pm$ 0.4 & 4.3 $\pm$ 0.7 & -10.9 $\pm$ 0.5 & 14.8 $\pm$ 1.1 & -15.2 $\pm$ 0.8 & 47.8 $\pm$ 1.8 \\ 
A6 $^\dag $ &  &   & -9.1 $\pm$ 0.5 & 29.3 $\pm$ 1.0 & -7.4 $\pm$ 0.5 & 28.4 $\pm$ 1.0 & -15.1 $\pm$ 0.8 & 52.5 $\pm$ 1.8 \\ 
Cav2 $^\dag $ &  &   & 21.3 $\pm$ 0.4 & 38.4 $\pm$ 0.7 & 17.2 $\pm$ 0.5 & 14.3 $\pm$ 1.1 & 12.8 $\pm$ 0.7 & 39.0 $\pm$ 1.8 \\ 
Cav3 $^\dag $ &  &   & 36.5 $\pm$ 0.4 & 56.4 $\pm$ 1.0 & 26.0 $\pm$ 0.6 & 40.3 $\pm$ 1.3 & 16.9 $\pm$ 0.7 & 36.8 $\pm$ 1.8 \\ 
Bow1 $^\dag $ &  &   & 56.0 $\pm$ 0.4 & 59.4 $\pm$ 1.6 & 36.4 $\pm$ 0.3 & 55.6 $\pm$ 0.8 & 23.4 $\pm$ 0.8 & 20.5 $\pm$ 1.6 \\ 
Bow2 $^\dag $ &  &   &  &   & 33.0 $\pm$ 0.5 & 42.8 $\pm$ 1.0 & 19.8 $\pm$ 0.8 & 58.8 $\pm$ 2.8 \\ 
In2 $^\dag $ &  &   & 21.5 $\pm$ 0.7 & 56.4 $\pm$ 1.3 & 10.9 $\pm$ 0.5 & 39.2 $\pm$ 1.0 & 12.5 $\pm$ 0.9 & 32.4 $\pm$ 2.0 \\ 
 \hline 
 \end{tabular} 
\\ \small  $\dag $ Regions where only MIRI data is available.\\
 \small   The $v_{peak}$ uncertainties are derived from the Gaussian fit.\\
 \small    the $\Delta v$ values are deconvolved from the instrumental resolution. 
 
\end{table*}

\begin{figure}[t]
    \centering
\includegraphics[width=0.5\textwidth,keepaspectratio]{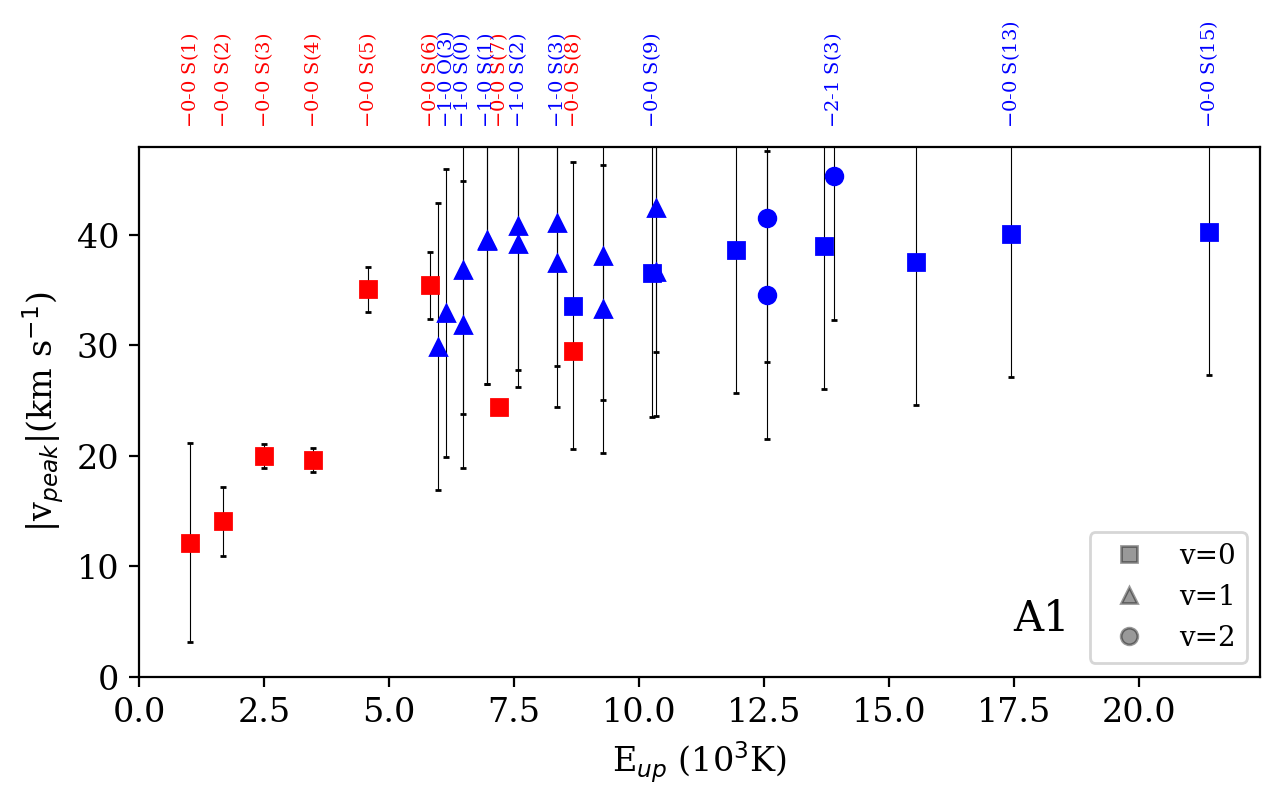}
    \caption{Absolute value for the radial velocity vs the energy of the level for knot A1. Red and blue symbols indicate MIRI and NIRSpec measurements. Different  vibrational levels are indicated with squares (v=0), triangles (v=1), and circles (v=2). The combined errors include those from Gaussian fits and instrumental uncertainties. }
    \label{fig:vel_energy}
\end{figure}

Given the observed radial velocity difference between the 2.12 \um\ and 8.03 \um\ \htwo\ transitions, we investigated whether a correlation exists between radial velocity and excitation energy. Table~\ref{tab:kinematics} lists the peak velocity and line width for representative transitions covering different excitation energies and outflow regions. 
A trend is apparent in which the radial velocity increases with excitation energy, with differences of up to $\sim$ 20-30 \kms\ between the lowest- and highest-excitation lines. This effect is most evident in knot A1, as illustrated in Figure~\ref{fig:vel_energy}, which shows radial velocity as a function of excitation energy. The associated uncertainties were computed as the quadrature sum of the Gaussian fitting errors and the absolute wavelength calibration uncertainty \citep{Argyriou2023}.

As shown in Figure~\ref{fig:vel_energy}, the velocity increases with excitation energy up to $\sim 5000$ K, above which it remains roughly constant, albeit with some fluctuations. In particular, the difference in velocity among lines with similar excitation energy gives an indication of the true calibration uncertainty . For example, 
the fact that the 0-0 S(7) radial velocity is sistematically lower than the 1-0 S(0)-S(2) velocities suggests that the calibration error of the MIRI channel 1 SHORT is larger than for the others MIRI channel.
The observed behavior is expected in lines tracing shock emission, as in the post-shocked regions the temperature and velocity decrease as the gas is cooled and slowed down \citep[e.g.][]{kaufman&neufeld1996}. Consequently, while the high-energy transitions probe the hotter gas close to the shock front, the low-energy lines trace post-shocked regions at low velocity and temperature. An alternative interpretation could be if the shock front is not planar but curved (i.e. a bow-shock) so progressively lower excitation lines probe progressively slower and more oblique shocks further from the bow-shock apex, in the bow shock wings.



\subsection{Comparison with shock models}

Molecular hydrogen can be excited either in photo-dissociation regions (PDRs) illuminated by UV photons or in non-dissociative shocks. Since the observed H$_2$ lines are spectrally resolved and their velocity peaks are mostly shifted from the systemic velocity, we infer that shock excitation provides a significant contribution to the observed emission. However, evidence for UV-irradiated gas has also been reported along the cavity walls from far-IR and sub-mm observations of CO, H$_2$O, and OH \citep{vankempen2009, vankempen10, Karska18}, suggesting that both excitation mechanisms likely coexist in the region.

To investigate the combined contribution of shocks and UV-induced excitation, we compared the observed \htwo\ emission with the models of \citet{kristensen2023}, which explore low-velocity shocks both with and without the influence of an external UV radiation field. Specifically, \citet{kristensen2023} developed a grid of one-dimensional stationary shock models using the Paris–Durham shock code \citep{flower2003}, incorporating a semi-isotropic external UV field \citep{godard2019}. This radiation field produces two main effects: it acts as an irradiated precursor that modifies the shock structure, and it enhances H$_2$ excitation through radiative pumping of electronic transitions followed by fluorescence.

In these models, the parameters defining the shock type are the pre-shock density ($n_H$), shock velocity ($v_s$), transverse magnetic field strength ($b$), UV radiation field strength ($G_0$), cosmic-ray ionization rate ($\zeta_{H_2}$), and the abundance of polycyclic aromatic hydrocarbons (PAHs, $X$(PAH)).  
Depending on the initial parameters, and in particular on the combination of shock velocity and magnetic field strength, the model solution is either a jump (J-) or a continuous (C-) shock or a combination of the two.

We compared the observed and modelled upper-level column densities and minimized the reduced $\chi^2$ to identify the model that best reproduces the extinction-corrected \htwo\ emission. The uncertainty on the observed column densities includes the statistical errors on the line fluxes, a calibration uncertainty of 5\%, and a typical uncertainty of 2 mag in the adopted extinction values.

The inclusion of transitions over an excitation range as large as possible ensures a much better constraint of the different parameters.
The models were normalized to the 0–0 S(1) data point, considering that filling factor effects or deviations from the assumed plane-parallel shock geometry may affect the column densities of individual transitions

To limit the parameter space, we fixed the PAH abundance and the cosmic ray ionization rate to the minimum values allowed by the grid, namely $X$(PAH)=$1.0 \times 10^{-8}$ and $\zeta_{H_2}=1.0 \times 10^{-17}$, respectively. This choice was motivated by the fact that we did not observe PAH emission in any outflow positions, indicating that their contribution to the outflow, if present, should be small. In addition, the contribution from cosmic ray ionization is expected to be very low, as no emission from ionised species is observed at the \htwo\ peaks. However, \cite{kristensen2023} shows that the models are only weakly dependent on the $\zeta_{H_2}$ parameter and become insensitive to it for UV fields $G_0 \ga 1$.

\begin{figure*}[ht]
    \centering
\includegraphics[width=1\textwidth,keepaspectratio]{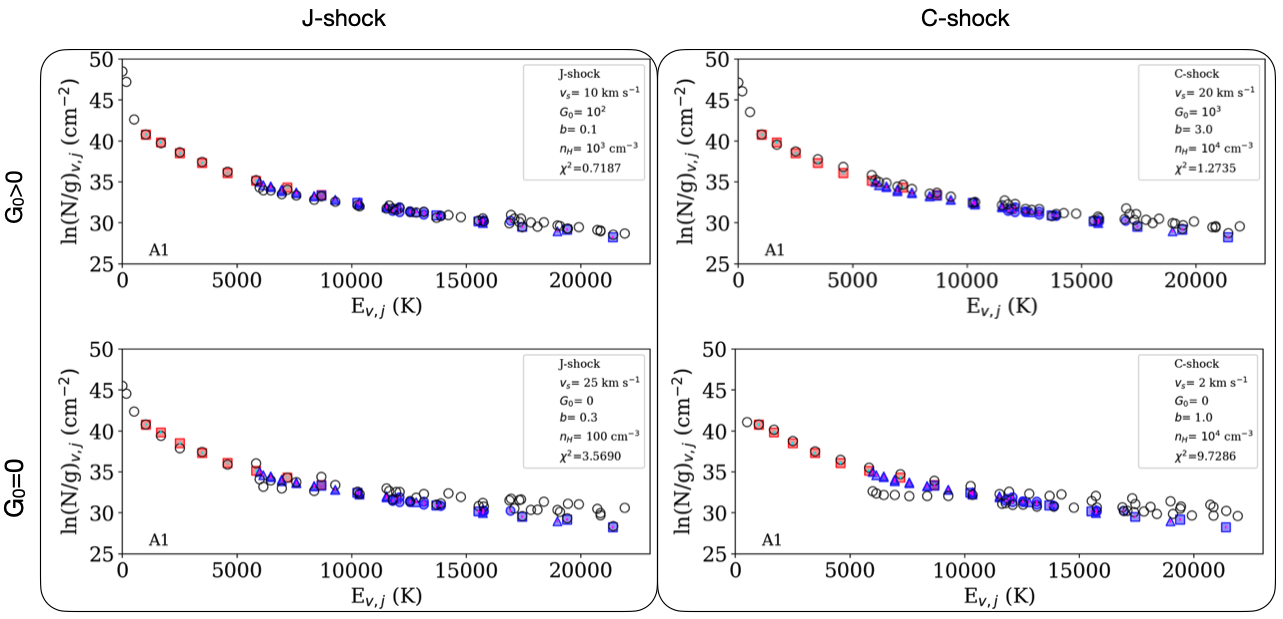}
\caption{
Best-fit shock models (black open circles) for knot~A1. 
The left and right panels correspond to the best J-type and C-type shock models, respectively, 
while the upper and lower panels show models with and without an external UV radiation field ($G_0$). 
Red and blue symbols represent MIRI and NIRSpec data, respectively, 
with different markers (squares, triangles, circles, and stars) denoting vibrational levels from $v=0$ to $v>2$.}
    \label{fig:shock_modelA1}
\end{figure*}

To better explore the parameter space, we separately fitted models with and without an external UV field, and for both C- and J-shocks. The results for knot A1 are presented in Figure \ref{fig:shock_modelA1}, where the best-fit model is displayed for each type of shock considered. 


From Figure \ref{fig:shock_modelA1}, it can be seen that the minimimum $\chi^2$ is attained for a J-type shock model with shock velocity 10 \kms , pre-shock density n$_{H}$= 10$^3$ \cmt, transverse magnetic field strength $b=0.1$ and UV field strength $G_0$ = 100. 
The C-shock best-fit model (case with $G_0 = 0$, with a $\chi^2$ a factor of two larger than for the best fit J-type model) overestimates the observed emission of both the high-energy and pure-rotational lines, while models of J- and C-type without the inclusion of the radiation field provide a significantly poorer match to both MIRI and NIRSpec observations ($\chi^2$ $\sim$ 5-15 larger than the best fit model).

To better evaluate the goodness of the best-fit models, we present in Figure \ref{fig:chisq} (Appendix) a $\chi^2$ corner map for knot A1, illustrating how the $\chi^2$ values vary and are minimized across the different free parameters of the models.
Each panel corresponds to a pair of parameters, with $\chi^2$ values minimized over all remaining variables.
From this figure, we infer that $G_0 >$ 100 and $v_s \sim$ 10-20 \kms\ are relatively well constrained, yielding $\chi^2$ values at least one order of magnitude smaller than for other parameter combinations.
Values of $b$ and $n_H$ in the ranges $0.1–3$ and $10^3-10^5$\cmt\ also provide acceptable fits, with $\chi^2 \la 3$ relative to the best solution.
However, solutions with $b >$ 0.3 and $n_H > 10^3$ \cmt\ can be ruled out, as the corresponding models predict \htwo\ emission widths larger than 1000 au for the 0–0 S(1) transition (i.e., $>$ 2\asec\ at the distance of HH 46), which would have been spatially resolved in our MIRI observations.

The results for all evaluated regions, along with their corresponding best-fit models, are presented in Figure \ref{fig:shockmodall} (Appendix) and summarized in Table \ref{tab:shockmod}.
In all cases, low-velocity ($v_s \sim$ 10 \kms) J-type shocks illuminated by a non-zero external UV field provide the best match to the observed \htwo\ emission.
The derived parameters are consistent across regions, with $G_0$ values between 10 and 100, magnetic field parameter $b \sim 0.1$, and pre-shock densities $n_H$ between $10^2$ and $10^3$ \cmt.
The results show good internal consistency: the strongest UV radiation field ($G_0$) is found toward knot A1, while the pre-shock density ($n_H$) is about one order of magnitude lower inside the cavity.
These findings are further discussed in Section \ref{sec:disc}.

\section{Discussion} \label{sec:disc}
\subsection{Physical conditions associated with the \htwo\ emission}

We derived the temperatures, column densities, visual extinction, and OPR in representative regions of the HH~46 outflow by fitting the H$_2$ rotational diagrams using two complementary approaches. The first assumes a two-temperature component fit, while the second considers a more realistic scenario in which the gas along the line of sight exhibits a continuous temperature stratification.

Our analysis shows that the linear fit underestimates the total H$_2$ column density by up to one order of magnitude, as it neglects the contribution from the colder gas component. In particular, fitting only the lowest rotational transitions yields temperatures of 
$T_{\text{warm}} \sim 600 - 850$ K and corresponding warm H$_2$ column densities $N_{\rm H_2-warm}$ of the order of 10$^{18}$ \cmd. In contrast, the temperature-stratified model reveals the presence of a substantial amount of colder gas ($T <$ 200 K) in localized regions, with total H$_2$ column densities $N_{\rm H_2}$ reaching $\sim$10$^{19}$ \cmd. Similarly, the linear fit underestimates the contribution from the hottest component.

Therefore, the temperature stratification approach provides a more physically realistic description of the H$_2$ emitting gas. Under this assumption, we infer OPR ratios close to the equilibrium value of three in most regions. Only in some redshifted areas, specifically within the cavity, the OPR values are lower, down to $\sim$1.2–1.5. Comparable or even lower OPR values have been derived from \textit{Spitzer} observations in several shock-excited outflows from very young protostars \citep[e.g.,][]{neufeld2006, neufeld2009, nisini2010}.  

If reactive collisions with atomic hydrogen are the dominant mechanism for para-to-ortho conversion \citep{neufeld2006}, the resulting OPR depends on the gas temperature, the duration of heating, and the atomic hydrogen density. This scenario is consistent with the lower OPR values typically found in Class 0 outflows, which have shorter dynamical timescales. The fact that we find near-equilibrium values over most of the flow suggests that the gas has been heated long enough to reach thermal equilibrium.  
Interestingly, OPR values below three are found in regions of the redshifted lobe where the lowest minimum temperatures ($T \sim 100$~K) are measured. At these lower temperatures, the para-to-ortho conversion proceeds more slowly, implying that the gas has not yet had sufficient time to reach equilibrium.

We have also constructed maps of the physical parameters that characterize the warm component traced by MIRI (Figure \ref{fig:allmaps}). The temperature map shows that the warmest regions correspond to active shock spots (knots, bow shocks, cavity walls). 
In particular, there is no evidence of a temperature stratification in the molecular emission from the axis to the border of the outflow, as it would be expected in MHD disk models and observed in other outflows \citep[e.g.][]{delabrosse2024,pascucci2025,caratti2024}.  
In the red-shifted lobe of the MIRI $T_{\text{warm}}$ map (see Figure \ref{fig:allmaps}), the warmest gas appears to be associated with the inner high velocity jet and the bow shock. A similar warm gas is also detected along the cavity walls. Instead, no association with the jet is seen in the temperature of the blue-shifted lobe, which reinforces the suggestion that the jet is fully atomic.

\subsubsection{Shock conditions}
Several pieces of evidence suggest that the primary excitation mechanism driving the \htwo\ emission is associated with shocks.
Most notably, the emission—aside from that tracing the main cavity walls—is concentrated in compact knots or arc-shaped features resembling bow shocks. Furthermore, the radial velocities ($v \sim$ 20-40 \kms ) and the resolved FWHM of the NIRSpec lines ($\Delta v$ $\ga$ 50 \kms) are consistent with supersonic gas motion and significant velocity dispersion, as expected in bow-shock regions.

Comparison with the shock model grid of \citet{kristensen2023} indicates that low-velocity J-type shocks, with pre-shock densities of $n_{H} = 10^2$–$10^3$\cmt\ and shock velocities of $v_s$ = 10 \kms, exposed to an external UV radiation field of $G_0 = 10$–$100$, provide the best match to the observed excitation diagrams in the regions covered by both the MIRI and NIRSpec observations. The inclusion of the near-IR lines observed by NIRSpec is crucial to break degeneracies among the model parameters \citep[e.g.,][]{valentin25, Vleugels25}.  
A better fit is obtained when a UV field enhanced by a factor of 10–100 relative to the diffuse interstellar radiation field is included.  
The most plausible origin of such a local FUV field is emission produced by the energetic shocks occurring within the jet itself, as evidenced by the detection of \neiii\ and \ariii\ lines along its beam, which indicate the presence of highly dissociative shocks (see Paper~I).
Another possible origin for the UV field is emission from the central source escaping through the evacuated cavity, consistent with the higher G$_0$ values derived near the source (knot A1).  
A less likely explanation is external UV irradiation impinging on the protostellar envelope; however, such radiation would be rapidly absorbed and would not penetrate deeply enough to affect the cavity interior.

In low-velocity J-shocks we would also expect to detect emission from atomic lines. In particular, in the considered range, the emission from the \si\ 25.2 \um\ line traces a medium where the ionisation is low. 
Among the considered regions, \si\ emission has been detected only in knots A1 and A2, where the observed \si/\htwo 17 \um\ ratio is $\sim 0.5$ consistent with the predictions of the best-fit shock model.
In the remaining regions, the model—assuming a lower pre-shock density of 100 \cmt—predicts \si\ emission to be an order of magnitude fainter than the \htwo\ 17 \um\ line, rendering it undetectable within the sensitivity limits of our observations.

We find shocks with shock velocity of $v_s \sim$ 10 \kms. 
The observed \htwo\ radial velocities range from $v$=0–20 \kms\ for the S(1)–S(4) lines and up to $v$=40 \kms\ for higher-energy transitions.  To reconcile these values, taking into consideration also inclination effects, we must assume that the shock interacts with a medium moving at a velocity of $v_m>$30 \kms. 

The derived shock models predict maximum post-shock densities of $n_{H,max}$ = 7$\times 10^3$ - 7$\times 10^4$ \cmt.  
As discussed in Section~\ref{sec:dfit}, the nearly complete thermalization observed between the rotational and ro-vibrational lines would require densities $> 10^5$ \cmt\ if the excitation were purely collisional.  
In the inferred shocks, however, thermalization is achieved at lower densities owing to the presence of a non-negligible radiation field.  
In irradiated shocks, both collisions and UV pumping contribute to the excitation of ro-vibrational levels, enhancing the thermalization process even for moderate UV field strengths (G$_0 \sim 10$–100; \citealt{godard2019}).

Finally, we note that the fitted shock models predict \htwo\ emission region sizes of $\la$50 au (i.e., $\sim$0.\asec1), which are below the spatial resolution of both NIRSpec and MIRI. Consequently, the post-shock cooling length is not expected to be resolved, consistent with the absence of any measurable spatial offset among lines of different excitation energies that trace distinct zones of the post-shock gas.

\subsection{On the origin of the \htwo\ emission structures}\label{sec:originh2}

In Paper~I, we discussed possible origins of the observed H$_2$ structures based on their emission morphology and the relative displacement between the atomic jet and the molecular outflow.
We suggested, in particular, that the observed morphological differences and the marked asymmetry in the blue-shifted region could arise from two distinct outflows driven by the two components of the HH 46 IRS binary system. High-resolution NIRCam images resolve the two sources, revealing an alignment between the atomic jet and the H$_2$ A1 bow shock with the respective members of the binary. Jet precession and environmental effects may further contribute to shaping the overall H$_2$ morphology within the blue-shifted cavity.

\begin{figure}[t]
    \centering
\includegraphics[width=0.5\textwidth,keepaspectratio]{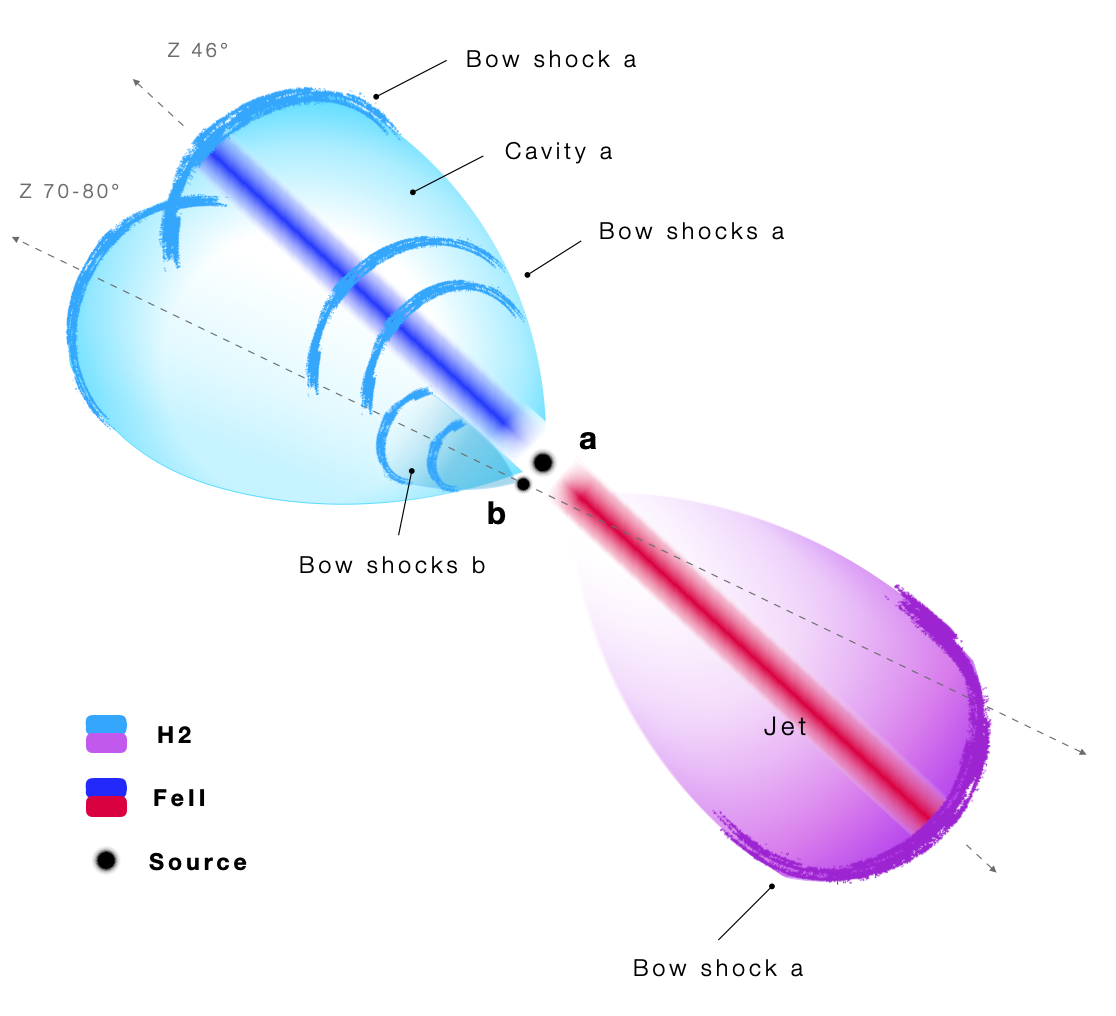}
    \caption{Schematic representation of the observed \htwo\ structures based on the emission morphology, analogous to Figures \ref{fig:nircam} and \ref{fig:regions}. The sources a$^\star$ and b$^\star$ are marked as black filled circles. The collimated atomic jet is illustrated in blue and red, while the \htwo emission is shown in cyan and violet. Regions of enhanced \htwo\ emission are indicated with more intense colors and are spatially connected to each component of the binary system. The figure is an artistic rendering by A. Spazzoli, designed specifically for this work.}
    \label{fig:ale}
\end{figure}

Our analysis provide support to the hypothesis of the presence of two separate outflows (see Figure \ref{fig:ale}). 
First, the NIRSpec H$_2$ line maps reveal an additional H$_2$ emission arc located outside the A1 structure, aligned with the direction of source~b$^\star$ and labeled as A1b.
In addition, our 2.12 \um\ radial velocity map, although obtained at a lower spectral resolution than the VLT/SINFONI observations of \citet{birney2024}, reveals clear and compelling spatial structures.
The regions at the highest blue-shifted velocities ($\sim -40$ \kms) are observed across the  A1 \htwo\ bow shock and the secondary newly detected arc A1b, further suggesting that these structures can be associated with two separate shock working surfaces traveling from source b$^\star$ with a position angle (PA) roughly between 70 and 80 degrees. 

The observed $\sim$20 \kms\ difference in the the \htwo\ 2.12\um radial velocities between knot A1 and the other H$_2$ emission peaks within the blue-shifted cavity may arise from a different inclination angle between the outflows driven by sources~a$^\star$ and~b$^\star$. Assuming that the outflow from source~a$^\star$ is inclined by $\sim$37° with respect to the plane of the sky—consistent with the inclination of the atomic jet—and that both outflows share similar total H$_2$ velocities (as suggested by the comparable shock velocities inferred for the various knots), the observed radial velocity offset implies that the outflow from source~b$^\star$ is inclined by approximately 55°.

We do not detect any collimated jet, either atomic or molecular, associated with the source b$^\star$ outflow, which would suggest that the observed arc-shaped structures are not jet-driven but originate from a wide-angle molecular outflow. 
However, we cannot exclude the presence of an unresolved compact jet as observed in similar outflows forming expanding bubbles, such as XZ Tau or SVS13 \citep{krist2008,Coffey04,hodapp2014}. 
Interestingly, there is no evidence of a counterpart of the outflow from source b$^\star$ in the red-shifted lobe along the same PA, suggesting that the secondary source produces a monopolar shell. 

The action of the collimated atomic jet from source~a$^\star$ is most evident in the red-shifted lobe, where the jet drives the gas outward, carving a cavity and producing distinct bow shocks upon interaction with denser regions. 
The more distant H$_2$ shells observed within the blue-shifted cavity (Figures~\ref{fig:regions} and~\ref{fig:ale}, Regions~A3–A6) display a kinematic structure at the MIRI resolution, suggesting that their origin is likely due to the combined action of the two outflows, which together compress and accelerate the surrounding medium.


In summary, the proposed scenario for HH46 — graphically illustrated in Figure \ref{fig:ale} — involves two outflows originating from a binary system: the primary component (source a$^\star$) drives a strong, collimated atomic jet, while the secondary (source b$^\star$) is responsible for a predominantly molecular outflow. This explains the marked differences between the red- and blue-shifted outflows, with the jet dominating the red-shifted structure, while the blue-shifted asymmetry reflects the coexistence of the primary's atomic jet and the secondary's monopolar wide-angle wind or unresolved compact jet with a different inclination angle. 

The H$_2$ emission, confined to the cavity walls and bow shocks, shows no significant morphological or collimation differences among lines of different excitation energies.
This indicates that the H$_2$ emission originates from shock interactions between jet-driven bow shocks and/or wide-angle winds with the ambient medium and the cavity walls. A similar behavior has been reported only for another Class~I source, Ced~110~IRS4 \citep{narang2025}, whereas other Class~0/I protostellar outflows studied with \textit{JWST} show a clear layered excitation structure (e.g., HH~211, \citealt{caratti2024}; TMC1-E, \citealt{Tychoniec2024}; HOPS~315, \citealt{vleugels2025}; L1448, Navarro et al., in prep.).


As shown in Paper I, CO (2-1) emission lies outside the \htwo\ emission, especially in the redshifted lobe. Similar layered structures among the hot atomic jet, warm \htwo\, and cold CO outflows have been interpreted in other objects as signatures of MHD disk winds \citep[e.g.][]{delabrosse2024,pascucci2025,caratti2024}. However, this layered structure is not observed among \htwo\ lines with different excitation conditions. We detect no clear temperature or velocity gradient in the \htwo\ emission, contrary to what is expected by disk-wind models \citep[e.g.][]{wang2019}. In fact, the radial velocity gradient traced by \htwo\ is the opposite of what is expected, with higher velocities at the cavity edges and lower velocities inside. This suggests that in HH46 the CO low-velocity flow results from gas swept up and entrained by shocks responsible for \htwo\ emission, rather than being directly ejected from a disk wind. 

To confirm the hypothesis that the \htwo\ bow-shock A1 is driven by source b$^\star$ we should measure its proper motion and check its expanding direction. The measured \htwo\ 1-0 S(1) 2.12 \um\ radial velocity in Knot A1 is $v \sim 40$ \kms. If we assume an inclination angle of $\sim$ 55 degrees (Section \ref{sec:originh2}), this would correspond to a tangential velocity of $v_{\text{tg}} \sim$ 28 \kms . At the NIRSpec spatial resolution (90 au at 450 pc), it would take about 7 years to observe the \htwo\ proper motion of the expanding shell. Considering that the presented observations were conducted in early 2023, follow-up NIRSpec observations could detect the outflow’s motion from year 2030 onwards.

\subsection{Mass loss and accretion rate of source b$^\star$}

Information about the accretion of the HH46 IRS binary has been given by \cite{Antoniucci08}, who measured a binary system's total mass accretion rate of $\dot M_{acc} = 2.2\times 10^{-7}$ \msunyr. 
Since mass accretion and ejection are intimately linked, indirect information on the accretion activity of the low-luminosity companion source b$^\star$ can now be inferred from the mass loss traced by the \htwo\ outflow knot A1, which is assumed to represent the most recent ejection episode from source b$^\star$.


We measure the mass loss rate $\dot M_{\text{loss}} (\text{H}_2)$ with two different methods. The first method computes the mass loss from the \htwo\ column density estimated on knot A1:

\begin{equation}
    \dot M_{\text{loss}} (\text{H}_2) = 2 \mu \, \text{m}_H \times \text{N}_{\text{H}_2} \text{A} \times (v_{\text{tg}}/dl)
\end{equation}

where $N_{\rm H_2}$ is the column density averaged over the emitting area $A$, $dl$ is the projected knot cross-section, and $v_{\text{tg}}$ is the knot tangential velocity.   
We consider the $N_{\rm H_2}$ total column density obtained from the stratification model, which takes into account also the contribution from the colder gas. For $A$ we assume a circular area of diameter $dl$= 0.\asec8 (i.e., $5.4\times 10^{15}$ cm at a distance of 450 pc) corresponding to $ A = 2.29\times 10^{31}$ cm$^2$. Measurements of the tangential velocity of \htwo\ knots close to the source are unavailable. 
Here we assume that the outflow traced by knot A1 has an inclination angle equal to 55 degrees, as estimated in Section \ref{sec:originh2}, and consequently derive $v_{\text{tg}}$ from the radial velocity 
measured on the S(2) 12 $\mu$m line, which is $v \sim$ 20 \kms. This leads to $v_{\text{tg}} \sim$ 14  \kms\ and $\dot M_{loss} (\text{H}_2) = 4.9\times 10^{-9}$ \msunyr. Variations of the outflow inclination angle by $\pm$10 degrees would vary the mass loss rate between $\dot M_{loss} (\text{H}_2) = 3.2 - 6.6 \times 10^{-9}$ \msunyr.  
The above $\dot M_{loss} $ determination is computed at the shock front, and should therefore be considered as un upper limit of the mass loss entering into the shock, due to the compression that increases the post-shocked density \citep[e.g.][]{Hartigan1994}.

Alternatively, the $\dot M_{loss} $ of the wind driving the shock can be estimated adopting the fitted shock parameters. In this case, we have: 

\begin{equation}
    \dot M_{\text{loss}} (\text{H}_2) =  \mu \,\text{m}_H  \text{n}_H \times {v_s} \times \text{A}
\end{equation} 

where $\text{n}_H$ and ${v_s}$ is the pre-shock density and shock velocity estimated for knot A1, i.e. 10$^3$ \cmt\ and 10 \kms , respectively, while $ \text{A}$ is the cross-section that can be assumed as the half-sphere surface of the shock front, with a radius equal to 0.\asec4. This leads to  
$\dot M_{loss} (\text{H}_2) = 2.8\times 10^{-11}$ \msunyr. 
If the shock is driven by a wind or jet having a velocity $v_{w}$, then the  $\dot M_{loss} $ has to be further corrected by a factor  ${v_w/v_s}$. For example, assuming $v_s \sim 100$ \kms\, implies $\dot M_{loss} (\text{jet/wind}) \sim 2.8\times 10^{-10}$ \msunyr,  i.e. about a factor of 10 lower than the estimate adopting the equation (1).

\cite{garcialopez2010} derived a mass-loss rate of $\dot M_{loss} (\text{jet}) = 0.5 - 2\times 10^{-7} $ \msunyr\ for the atomic jet driven by source a$^\star$, which is more than a factor of 200 higher than our estimate of the mass-loss rate for source b$^\star$. Assuming that the mass-loss rate scales proportionally with the mass-accretion rate, we infer that the primary source is accreting at least $\sim$200 times higher than the secondary. Given a total accretion rate of $\dot{M}{\mathrm{acc}} = 2.2 \times 10^{-7}$~\msunyr, we estimate that t$\dot{M}{\mathrm{acc}}$ for sourceb$^\star$ is lower than $10^{-9}$\msunyr.

\section{Conclusions}\label{sec:concl}   
In this article, we present \jwst\ observations of molecular hydrogen emission in the HH46 IRS outflow, obtained with the NIRSpec (IFU) and MIRI (MRS) instruments. Our analysis focuses on the morphology, kinematics, and excitation conditions of the detected \htwo\ emission, with the goal of investigating its origin and its connection to the collimated atomic jet.

Our conclusions can be summarised as follows:

\begin{itemize}

\item The \htwo\ pure rotational emission map observed with MIRI shows a complex morphology, with a marked asymmetry between the blue- and the red-shifted outflows. Peaks of emission are localised along the outflow borders, in compact shock spots not-aligned with the atomic jet, and in extended bow-shocks, pushed by the jet and/or by a wide-angle wind. The NIRSpec map of ro-vibrational emission close to the central source reveals, in particular, the presence of multiple shell structures, misaligned with the jet and  likely driven by the less luminous secondary source of the HH46 IRS binary system. This latter hypothesis is further confirmed by the kinematical analysis, showing a clear different velocity pattern in these shells with respect to the rest of the \htwo\ emission in the region.

\item  The \htwo\ rotational diagrams, constructed by combining both MIRI and NIRSpec data, indicate that the observed \htwo\ emission is well reproduced by a multi-temperature gas component, with temperatures ranging from $\sim$500 to 2000 K. 
A temperature stratification model provides a better fit to the column densities and reveals that the dominant contribution to the total \htwo\ column density arises from colder gas components, with temperatures down to 100 K, which are not traced by the H$_2$ lines observed with MIRI.
The \htwo\ OPR is generally found to be close to the equilibrium value, except in a few regions within the red-shifted lobe where values as low as 1.2 are inferred. These regions are characterized by lower minimum temperatures, suggesting that the gas has not yet had sufficient time to reach thermal equilibrium.

\item  We have constructed maps of temperature, extinction, column density and radial velocity that illustrate the excitation and kinematical pattern of the emission. In particular,  we show that the warmer gas is located at the edges of the flow, on individual \htwo\ knots and extended bow-shocks. We do not observe any temperature and radial velocity stratification in the \htwo\ emission within the cavity carved by the outflow, at variance with the expected behaviour predicted by magnetohydrodynamic (MHD) disk wind models. In the red-shifted outflow, the morphology, excitation and kinematics of the \htwo\ lines are clearly compatible with bow-shocks pushed by the atomic jet.

\item Although we do not observe morphological differences among lines tracing different excitation temperatures, we do observe radial velocity variations up to $\sim$ 20 \kms\ between lines with different upper energies. We interpret this finding assuming that the various lines trace not-resolved post-shocked regions with different velocities and temperatures as the gas cools down and it is gradually decelerated. 

\item We estimate that the mass loss rate of the molecular outflow from source b$^\star$ is $\sim$ 5$\times 10^{-10}$ \msunyr, thus more than two order of magnitudes smaller than the mass loss rate of the collimated atomic jet. This finding suggests that the secondary source should possess a mass accretion rate $\la$ 10$^{-9}$ \msunyr.

\item Comparisons with grids of shock models of \cite{kristensen2023} indicate that the \htwo\ excitation is compatible with low-velocity (10 \kms ) J-shocks with pre-shock densities of the order of 100-1000 \cmt. Irradiation from an external UV field with strength of $\sim$ 10-100 G$_o$ is needed to reproduce both the rotational and ro-vibrational line intensities observed with MIRI and NIRSpec. We suggest that such a field is produced in high velocity dissociative shocks occurring along the collimated jet.

\end{itemize}
The large spectral coverage and high spatial resolution of \jwst, combined with its adequate spectral resolution to infer kinematical information, demonstrate its unique capability for detailed studies of the origin and excitation of molecular hydrogen in outflows from young stars. Further comparison with similar studies conducted on a larger sample of YSOs of different ages and properties will shed light on how the conclusions derived for HH46 can be considered representative of similar outflows from Class I sources. 

\clearpage
\section*{Acknowledgments}


This work is based on observations made with the NASA/ESA/CSA James Webb Space Telescope. Data were obtained from the Mikulski Archive for Space Telescopes (MAST) at the Space Telescope Science Institute, which is operated by the Association of Universities for Research in Astronomy, Inc., under NASA contract NAS 5$-$03127 for \jwst. The specific observations associated with program PID1706 and can be accessed via \dataset[DOI:10.17909/eav 1-0619]{https://doi.org/10.17909/eav1-0619}.
NIRCAM Data used in Figure 1 from DDT program PID4441 (P.I. K. Pontoppidan) can be accessed via \dataset[DOI:10.17909/4qkr-a057]{https://doi.org/10.17909/4qkr-a057}.
We thank A. Spazzoli for the original artwork illustrating the HH46 outflow structure shown in Figure \ref{fig:ale}.
MGN thanks the European Union - NextGenerationEU, M4C2 1.2 CUP C83C25000450006. PH and HA acknowledge funding support from \jwst\ GO program \#1706 provided by NASA through a grant from the Space Telescope Science Institute. 
We gratefully acknowledges the help of the Space Telescope Science Institute \jwst\ Helpdesk, and in particular Jane Morrison, for her valuable suggestions on the NIRSpec data reduction. 
INAF co-authors acknowledge support from the Large Grant INAF 2022 “YSOs Outflows, Disks and Accretion: towards a global framework for the evolution of planet forming systems (YODA)” and from PRIN-MUR 2022 20228JPA3A “The path to star and planet formation in the \jwst\ era (PATH)”, funded by NextGeneration EU and from INAF-GoG 2022 “NIR-dark Accretion Outbursts in Massive Young stellar objects (NAOMY)”. P.J.K. acknowledges financial support from the Science Foundation Ireland/Irish Research Council Pathway programme under Grant Number 21/PATH-S/9360.
LP acknowledges financial support under the National Recovery and Resilience Plan (NRRP), Mission 4, Component 2, Investment 1.1, Call for tender No. 104 published on 2.2.2022 by
the Italian Ministry of University and Research (MUR), funded by the European Union – NextGenerationEU-Project Title 2022JC2Y93 Chemical Origins:
linking the fossil composition of the Solar System with the chemistry of protoplanetary disks – CUP J53D23001600006 – Grant Assignment Decree No.
962 adopted on 30.06.2023 by the Italian Ministry of Ministry of University and Research (MUR). 
LP also acknowledges the PRIN-MUR 2020 BEYOND-2p (Astrochemistry beyond the second period elements,
Prot. 2020AFB3FX), the project ASI-Astrobiologia 2023 MIGLIORA (Modeling Chemical
Complexity, F83C23000800005), and the INAF fundings GO 2024 ICES, GO 2023 PROTO-SKA, and Mini Grant 2022 Chemical Origins.
EvD acknowledges the funding from the European Research Council (ERC) under the European Union’s Horizon 2020 research and innovation programme (grant agreement No. 291141 MOLDISK). 
SC gratefully acknowledges support from the Programme National de Physique et Chimie du Milieu Interstellaire (PCMI), cofunded by CNRS-INSU, CNES, and CEA, and from Observatoire de Paris (Action Fédératrice Incitative Univers Froid).  
Views and opinions expressed are however those of the author(s) only and do not necessarily reflect those of the European Union or the European Research Council Executive Agency. Neither the European Union nor the granting authority can be held responsible for them. 


%

\vspace{5mm}
\facilities{\jwst(STIS)}






\clearpage

\appendix

\section{Extracted spectrum and detected \htwo\ lines in the outflow}
\begin{figure}[ht]
    \centering
\includegraphics[width=0.8\textwidth,keepaspectratio]{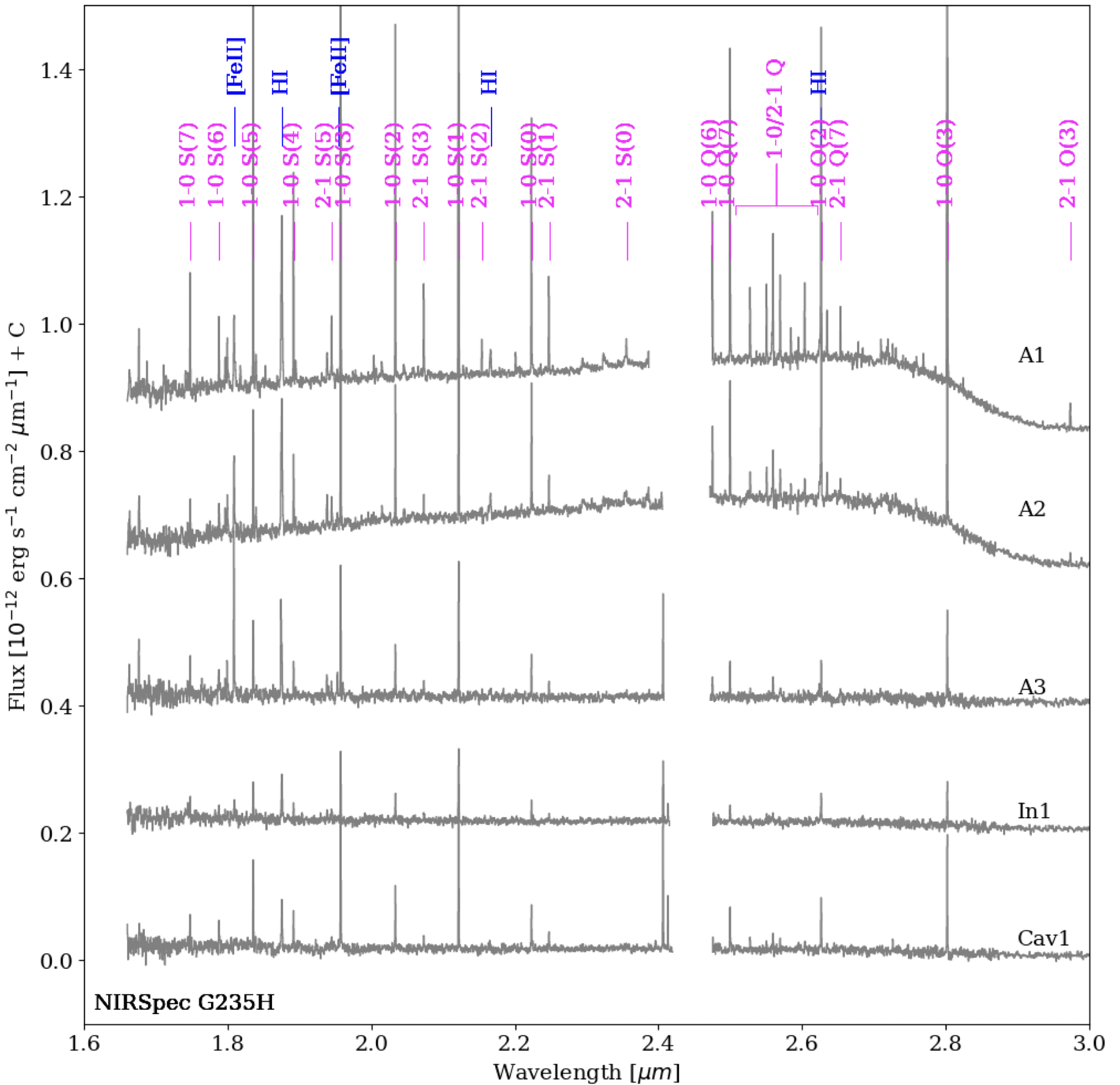}
    \caption{NIRSpec/G235H spectra extracted for the five representative regions located in the redshifted cavity, using a circular aperture with a radius of 0.\asec4. The main detected atomic (blue) and \htwo\ molecular (magenta) emission lines are indicated. The discontinuity visible between 2.4 \um\ and 2.5 \um\ corresponds to the detector gap of NIRSpec.
}
    \label{fig:NIRSPEC_specs1}
\end{figure}

\begin{figure}[ht]
    \centering
\includegraphics[width=0.8\textwidth,keepaspectratio]{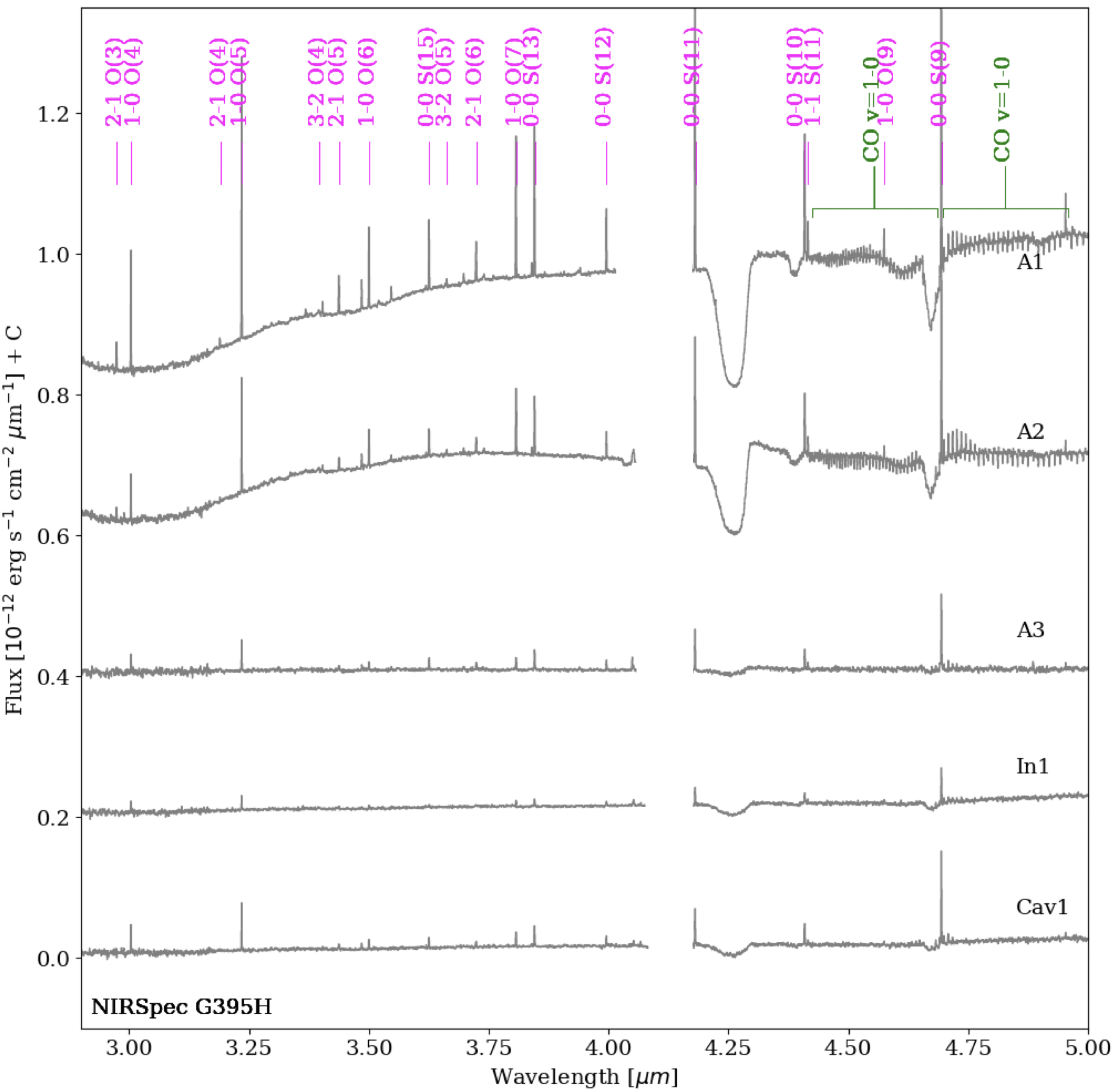}
    \caption{NIRSpec/G395H spectra extracted for the five representative regions located in the redshifted cavity, using a circular aperture with a radius of 0.\asec4. The main detected 
    molecular \htwo\ (magenta) and CO (green) lines are indicated. The discontinuity visible between 4.1 \um\ and 4.2 \um\ corresponds to the detector gap of NIRSpec.}
    \label{fig:NIRSPEC_specs2}
\end{figure}

\begin{figure}[ht]
    \centering
\includegraphics[width=1\textwidth,keepaspectratio]{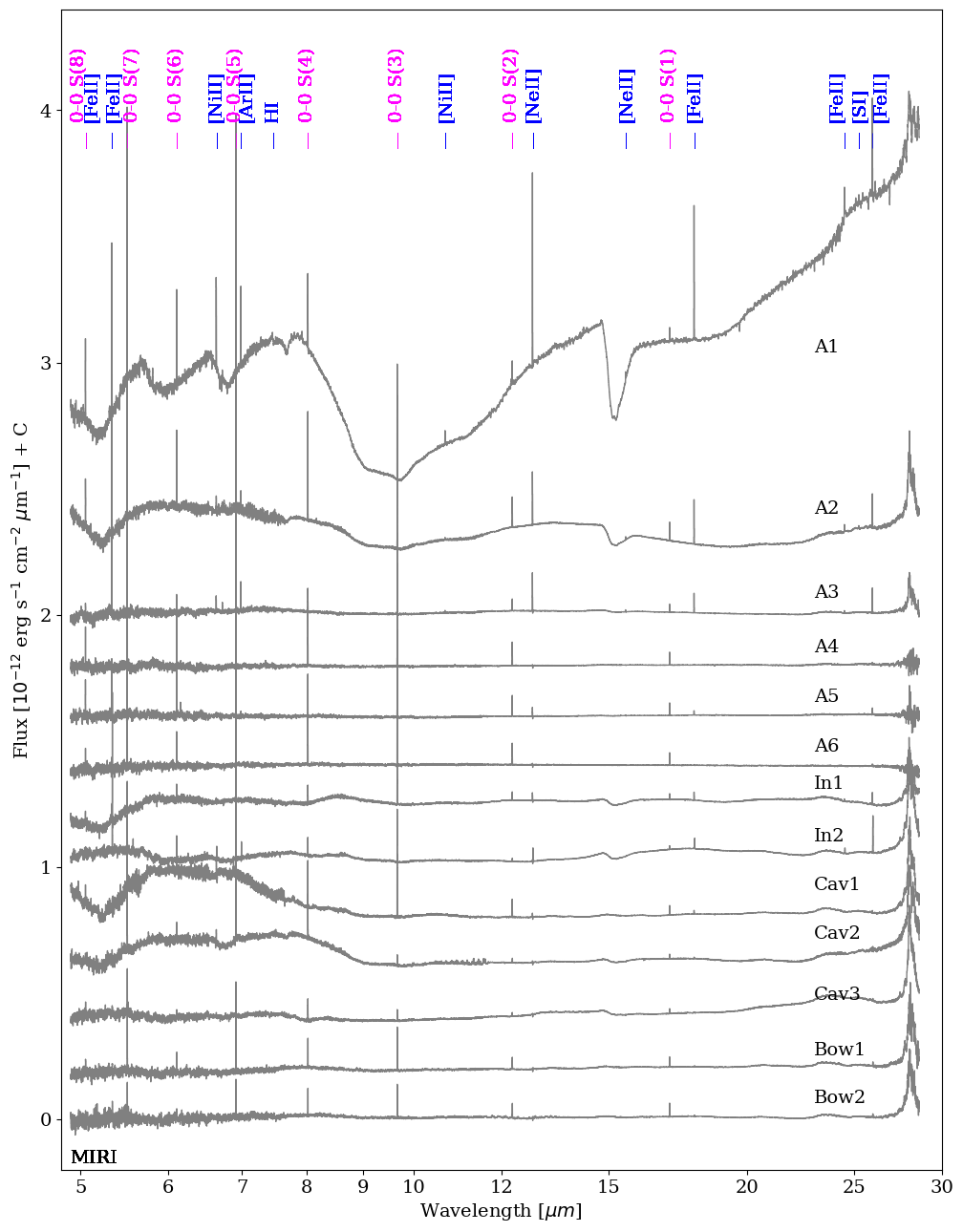}
    \caption{
    MIRI spectra extracted for the 13 representative regions located in both the red- and blueshifted cavities, using a circular aperture with a radius of 0.\asec4. The main detected atomic (blue) and H$_2$ molecular (magenta) emission lines are indicated.
    }
    \label{fig:MIRI_specs}
\end{figure}

Figures \ref{fig:NIRSPEC_specs1} and \ref{fig:NIRSPEC_specs2} display the NIRSpec spectra extracted for five representative regions (A1, A2, A3, Cav1, and In1) located within the blueshifted cavity, obtained with the G235H and G395H gratings, respectively. Figure \ref{fig:MIRI_specs} shows the spectra extracted from regions covered by MIRI, including both the redshifted and blueshifted portions of the outflow (A1, A2, A3, A4, A5, A6, Cav1, Cav2, Cav3, Bow1, Bow2, In1, and In2). All spectra were extracted using a circular aperture with a radius of 0.\asec4, centered on the positions listed in Table \ref{table:reg}. The detected \htwo\ lines and their corresponding fluxes are reported in Tables \ref{tab:nirspec_lines} and \ref{tab:miri_lines} for the NIRSpec and MIRI datasets, respectively.

\begin{center} 
 \begin{longtable*}{l l l l l l l} 
 \caption{List of detected NIRSpec and MIRI H2 lines in the representative regions (A1, A2, A3, Cav1 and In1).} 
 \label{tab:nirspec_lines} \\ \hline\hline 
   &   & A1 & A2 & A3 & Cav1 & In1 \\  
 Line ID &  $\lambda_{vac}$ & $F \pm \Delta F$ & $F \pm \Delta F$ & $F \pm \Delta F$ & $F \pm \Delta F$ & $F \pm \Delta F$ \\  
 & \multicolumn{1}{c}{\tiny ($\mu m$)}  & \multicolumn{5}{c}{ {\tiny ($10^{-13} erg/s/cm^2$)}}    \\ 
 \hline 
 \endfirsthead 
 \multicolumn{7}{c}%
 {{\bfseries \tablename\ \thetable{} -- continued from previous page}} \\ \hline\hline 
   &   & A1 & A2 & A3 & Cav1 & In1 \\  
 Line ID &  $\lambda_{vac}$ & $F \pm \Delta F$ & $F \pm \Delta F$ & $F \pm \Delta F$ & $F \pm \Delta F$ & $F \pm \Delta F$ \\  
 & \multicolumn{1}{c}{\tiny ($\mu m$)}  & \multicolumn{5}{c}{ {\tiny ($10^{-13} erg/s/cm^2$)}}    \\ 
 \hline 
 \endhead 
 \hline \multicolumn{7}{|r|}{{Continued on next page}} \\ \hline 
 \endfoot \hline \hline \endlastfoot 
 1.748 & 1-0S(7) & 1.96 $\pm$ 0.15 & 0.61 $\pm$ 0.09 & 0.47 $\pm$ 0.08 \\ 
1.788 & 1-0S(6) & 1.30 $\pm$ 0.08 & 0.56 $\pm$ 0.07 & 0.42 $\pm$ 0.06 \\ 
1.836 & 1-0S(5) & 5.99 $\pm$ 0.13 & 2.18 $\pm$ 0.08 & 1.20 $\pm$ 0.14 & 1.15 $\pm$ 0.08 & 0.44 $\pm$ 0.04 \\ 
1.853 & 2-1S(7) & 0.30 $\pm$ 0.06 &  &  &  \\ 
1.892 & 1-0S(4) & 3.28 $\pm$ 0.10 & 1.03 $\pm$ 0.06 & 0.60 $\pm$ 0.08 \\ 
1.895 & 2-1S(6) & 0.34 $\pm$ 0.04 &  &  &  &  \\ 
1.945 & 2-1S(5) & 1.11 $\pm$ 0.07 & 0.69 $\pm$ 0.09 &  &  \\ 
1.958 & 1-0S(3) & 13.51 $\pm$ 0.09 & 5.09 $\pm$ 0.07 & 1.58 $\pm$ 0.13 & 2.24 $\pm$ 0.07 & 1.04 $\pm$ 0.06 \\ 
2.034 & 1-0S(2) & 5.47 $\pm$ 0.05 & 2.26 $\pm$ 0.06 & 0.84 $\pm$ 0.08 & 0.93 $\pm$ 0.06 & 0.32 $\pm$ 0.04 \\ 
2.073 & 2-1S(3) & 1.70 $\pm$ 0.05 & 0.39 $\pm$ 0.05 &  & 0.24 $\pm$ 0.04 & 0.47 $\pm$ 0.06 \\ 
2.122 & 1-0S(1) & 16.15 $\pm$ 0.06 & 6.96 $\pm$ 0.06 & 2.12 $\pm$ 0.08 & 2.43 $\pm$ 0.05 & 1.23 $\pm$ 0.06 \\ 
2.154 & 2-1S(2) & 0.53 $\pm$ 0.04 &  &  &  \\ 
2.201 & 3-2S(3) & 0.23 $\pm$ 0.04 &  &  &  \\ 
2.223 & 1-0S(0) & 4.26 $\pm$ 0.05 & 1.94 $\pm$ 0.05 & 0.74 $\pm$ 0.06 & 0.58 $\pm$ 0.04 & 0.23 $\pm$ 0.04 \\ 
2.248 & 2-1S(1) & 1.53 $\pm$ 0.05 & 0.51 $\pm$ 0.04 & 0.28 $\pm$ 0.04 & 0.38 $\pm$ 0.05 \\ 
2.356 & 2-1S(0) & 0.49 $\pm$ 0.05 &  &  &  &  \\ 
2.476 & 1-0Q(6) & 2.09 $\pm$ 0.06 & 0.97 $\pm$ 0.07 &  &  \\ 
2.500 & 1-0Q(7) & 4.97 $\pm$ 0.06 &  & 0.70 $\pm$ 0.08 & 0.65 $\pm$ 0.05 & 0.32 $\pm$ 0.04 \\ 
2.528 & 1-0Q(8) & 1.10 $\pm$ 0.06 & 0.52 $\pm$ 0.06 &  \\ 
2.551 & 2-1Q(1) &  & 0.50 $\pm$ 0.06 &  &  \\ 
2.560 & 1-0Q(9) &  & 0.74 $\pm$ 0.08 &  \\ 
2.570 & 2-1Q(3) & 1.29 $\pm$ 0.05 & 0.39 $\pm$ 0.05 & 0.19 $\pm$ 0.03 &  \\ 
2.585 & 2-1Q(4) & 0.53 $\pm$ 0.06 &  &  &  \\ 
2.604 & 2-1Q(5) & 1.13 $\pm$ 0.05 & 0.47 $\pm$ 0.09 &  &  \\ 
2.627 & 1-0O(2) & 4.93 $\pm$ 0.09 & 2.72 $\pm$ 0.07 & 0.70 $\pm$ 0.11 & 0.72 $\pm$ 0.06 & 0.56 $\pm$ 0.08 \\ 
2.635 & 1-0Q(11) & 0.72 $\pm$ 0.05 & 0.33 $\pm$ 0.05 &  &  \\ 
2.654 & 2-1Q(7) & 0.70 $\pm$ 0.06 &  &  \\ 
2.803 & 1-0O(3) & 13.50 $\pm$ 0.05 & 6.74 $\pm$ 0.06 & 1.27 $\pm$ 0.11 & 1.62 $\pm$ 0.07 & 0.73 $\pm$ 0.07 \\ 
2.974 & 2-1O(3) & 0.49 $\pm$ 0.05 &  &  \\ 
3.004 & 1-0O(4) & 2.14 $\pm$ 0.07 & 0.88 $\pm$ 0.07 & 0.37 $\pm$ 0.05 &  \\ 
3.190 & 2-1O(4) & 0.22 $\pm$ 0.02 & 0.10 $\pm$ 0.02 &  \\ 
3.235 & 1-0O(5) & 6.41 $\pm$ 0.03 & 2.75 $\pm$ 0.04 & 0.78 $\pm$ 0.04 & 0.88 $\pm$ 0.02 & 0.35 $\pm$ 0.03 \\ 
3.438 & 2-1O(5) & 0.95 $\pm$ 0.03 & 0.28 $\pm$ 0.02 & 0.16 $\pm$ 0.03 &  \\ 
3.501 & 1-0O(6) & 1.76 $\pm$ 0.02 & 0.76 $\pm$ 0.03 & 0.23 $\pm$ 0.04 & 0.27 $\pm$ 0.02 & 0.16 $\pm$ 0.02 \\ 
3.626 & 0-0S(15) & 1.49 $\pm$ 0.03 & 0.61 $\pm$ 0.02 & 0.25 $\pm$ 0.03 & 0.22 $\pm$ 0.03 \\ 
3.663 & 3-2O(5) & 0.13 $\pm$ 0.02 &  &  &  \\ 
3.724 & 0-0S(14) & 0.98 $\pm$ 0.02 & 0.37 $\pm$ 0.03 &  \\ 
3.807 & 1-0O(7) & 3.01 $\pm$ 0.03 & 1.28 $\pm$ 0.02 & 0.26 $\pm$ 0.02 & 0.26 $\pm$ 0.02 & 0.15 $\pm$ 0.03 \\ 
3.846 & 0-0S(13) & 3.08 $\pm$ 0.09 & 1.23 $\pm$ 0.05 & 0.46 $\pm$ 0.03 & 0.48 $\pm$ 0.03 & 0.17 $\pm$ 0.02 \\ 
3.996 & 0-0S(12) & 1.46 $\pm$ 0.02 & 0.55 $\pm$ 0.02 & 0.22 $\pm$ 0.04 & 0.23 $\pm$ 0.03 \\ 
4.181 & 0-0S(11) & 6.23 $\pm$ 0.04 & 2.50 $\pm$ 0.07 & 0.85 $\pm$ 0.04 & 0.85 $\pm$ 0.05 & 0.51 $\pm$ 0.04 \\ 
4.410 & 0-0S(10) & 2.97 $\pm$ 0.08 & 1.29 $\pm$ 0.11 & 0.43 $\pm$ 0.05 & 0.49 $\pm$ 0.05 & 0.25 $\pm$ 0.02 \\ 
4.417 & 1-1S(11) & 0.74 $\pm$ 0.07 & 0.27 $\pm$ 0.05 & 0.10 $\pm$ 0.02 \\ 
4.575 & 1-0O(9) & 0.77 $\pm$ 0.06 &  &  \\ 
4.695 & 0-0S(9) & 11.35 $\pm$ 0.30 & 5.35 $\pm$ 0.17 & 1.71 $\pm$ 0.05 & 1.88 $\pm$ 0.09 & 0.90 $\pm$ 0.04 \\ 
4.954 & 1-1S(9) & 0.85 $\pm$ 0.05 & 0.25 $\pm$ 0.05 &  &  \\  \hline 
5.053 & 0-0S(8) & 5.28 $\pm$ 0.06 & 3.02 $\pm$ 0.05 & 0.84 $\pm$ 0.08 & 1.03 $\pm$ 0.04 &  \\ 
5.053 & 0-0S(8) & 5.72 $\pm$ 0.30 & 3.39 $\pm$ 0.19 & 1.66 $\pm$ 0.27 \\ 
5.511 & 0-0S(7) & 22.12 $\pm$ 0.27 & 14.80 $\pm$ 0.17 & 3.72 $\pm$ 0.20 & 5.45 $\pm$ 0.37 & 2.09 $\pm$ 0.14 \\ 
6.109 & 0-0S(6) & 7.36 $\pm$ 0.24 & 6.17 $\pm$ 0.16 & 1.32 $\pm$ 0.11 & 2.87 $\pm$ 0.21 & 1.08 $\pm$ 0.14 \\ 
6.910 & 0-0S(5) & 24.19 $\pm$ 0.24 & 24.19 $\pm$ 0.22 & 5.81 $\pm$ 0.08 & 11.35 $\pm$ 0.27 & 4.19 $\pm$ 0.09 \\ 
8.025 & 0-0S(4) & 8.25 $\pm$ 0.38 & 10.73 $\pm$ 0.08 & 2.36 $\pm$ 0.07 & 5.09 $\pm$ 0.10 & 1.80 $\pm$ 0.10 \\ 
9.665 & 0-0S(3) & 13.18 $\pm$ 0.08 & 17.40 $\pm$ 0.08 & 5.91 $\pm$ 0.07 & 12.71 $\pm$ 0.08 & 4.81 $\pm$ 0.05 \\ 
12.279 & 0-0S(2) & 3.95 $\pm$ 0.11 & 5.03 $\pm$ 0.03 & 2.00 $\pm$ 0.01 & 3.13 $\pm$ 0.08 & 1.48 $\pm$ 0.04 \\ 
17.035 & 0-0S(1) & 3.49 $\pm$ 0.21 & 4.23 $\pm$ 0.04 & 1.88 $\pm$ 0.01 & 2.34 $\pm$ 0.01 & 1.36 $\pm$ 0.03 \\ 
\end{longtable*} 
 \end{center}

 \begin{sidewaystable} 
 \centering 
 \caption{List of detected MIRI \htwo lines in the representative regions (A4, A5, A6, Cav2, Cav3, Bow1, Bow2 and In2) where NIRSpec data is not available.} 
 \label{tab:miri_lines} 
  \begin{tabular}{c c c c c c c c c c c c c c c c c c c c c c c} 
 \hline\hline 
 & & A4 & A5 & A6 & Cav2 & Cav3 & Bow1 & Bow2 & In2 \\
 Line ID &  $\lambda_{vac}$ & $F \pm \Delta F$ & $F \pm \Delta F$ & $F \pm \Delta F$ & $F \pm \Delta F$ & $F \pm \Delta F$ & $F \pm \Delta F$ & $F \pm \Delta F$ & $F \pm \Delta F$  \\  
 & \multicolumn{1}{c}{\small ($\mu m$)}  & \multicolumn{8}{c}{ {\small ($10^{-13} erg/s/cm^2$)}}    \\ 
 \hline 
 5.053 & 0-0S(8) & 2.40 $\pm$ 0.15 & 2.35 $\pm$ 0.14 & 1.64 $\pm$ 0.24 & 0.68 $\pm$ 0.12 \\ 
5.511 & 0-0S(7) & 11.12 $\pm$ 0.17 & 10.07 $\pm$ 0.17 & 6.66 $\pm$ 0.13 & 4.92 $\pm$ 0.17 & 3.24 $\pm$ 0.18 & 5.30 $\pm$ 0.22 & 2.55 $\pm$ 0.28 & 1.85 $\pm$ 0.18 \\ 
6.109 & 0-0S(6) & 4.44 $\pm$ 0.11 & 3.90 $\pm$ 0.17 & 3.08 $\pm$ 0.17 & 1.31 $\pm$ 0.17 & 1.98 $\pm$ 0.16 \\ 
6.910 & 0-0S(5) & 16.49 $\pm$ 0.10 & 12.06 $\pm$ 0.10 & 10.58 $\pm$ 0.09 & 7.05 $\pm$ 0.15 & 3.16 $\pm$ 0.09 & 6.13 $\pm$ 0.13 & 4.05 $\pm$ 0.14 & 2.42 $\pm$ 0.09 \\ 
8.025 & 0-0S(4) & 5.78 $\pm$ 0.07 & 4.11 $\pm$ 0.07 & 4.53 $\pm$ 0.07 & 5.45 $\pm$ 0.13 & 2.40 $\pm$ 0.12 & 3.44 $\pm$ 0.08 & 3.08 $\pm$ 0.07 & 2.22 $\pm$ 0.09 \\ 
9.665 & 0-0S(3) & 13.48 $\pm$ 0.04 & 10.04 $\pm$ 0.06 & 11.26 $\pm$ 0.06 & 1.69 $\pm$ 0.06 & 1.68 $\pm$ 0.04 & 4.93 $\pm$ 0.05 & 4.06 $\pm$ 0.07 & 1.08 $\pm$ 0.04 \\ 
12.279 & 0-0S(2) & 3.80 $\pm$ 0.02 & 3.43 $\pm$ 0.01 & 3.48 $\pm$ 0.03 & 1.00 $\pm$ 0.02 & 0.84 $\pm$ 0.03 & 2.35 $\pm$ 0.02 & 3.09 $\pm$ 0.21 & 0.70 $\pm$ 0.02 \\ 
17.035 & 0-0S(1) & 3.13 $\pm$ 0.01 & 3.07 $\pm$ 0.01 & 2.96 $\pm$ 0.02 & 1.09 $\pm$ 0.04 & 1.17 $\pm$ 0.03 & 2.56 $\pm$ 0.02 & 3.30 $\pm$ 0.02 & 0.85 $\pm$ 0.03 \\ 
 \hline 
 \hline 
\end{tabular} 
 \end{sidewaystable} 

\section{Temperature stratification for rotational diagrams of representative regions}
\begin{figure}[ht]
    \centering
chi schhccc\includegraphics[width=0.5\textwidth,keepaspectratio]{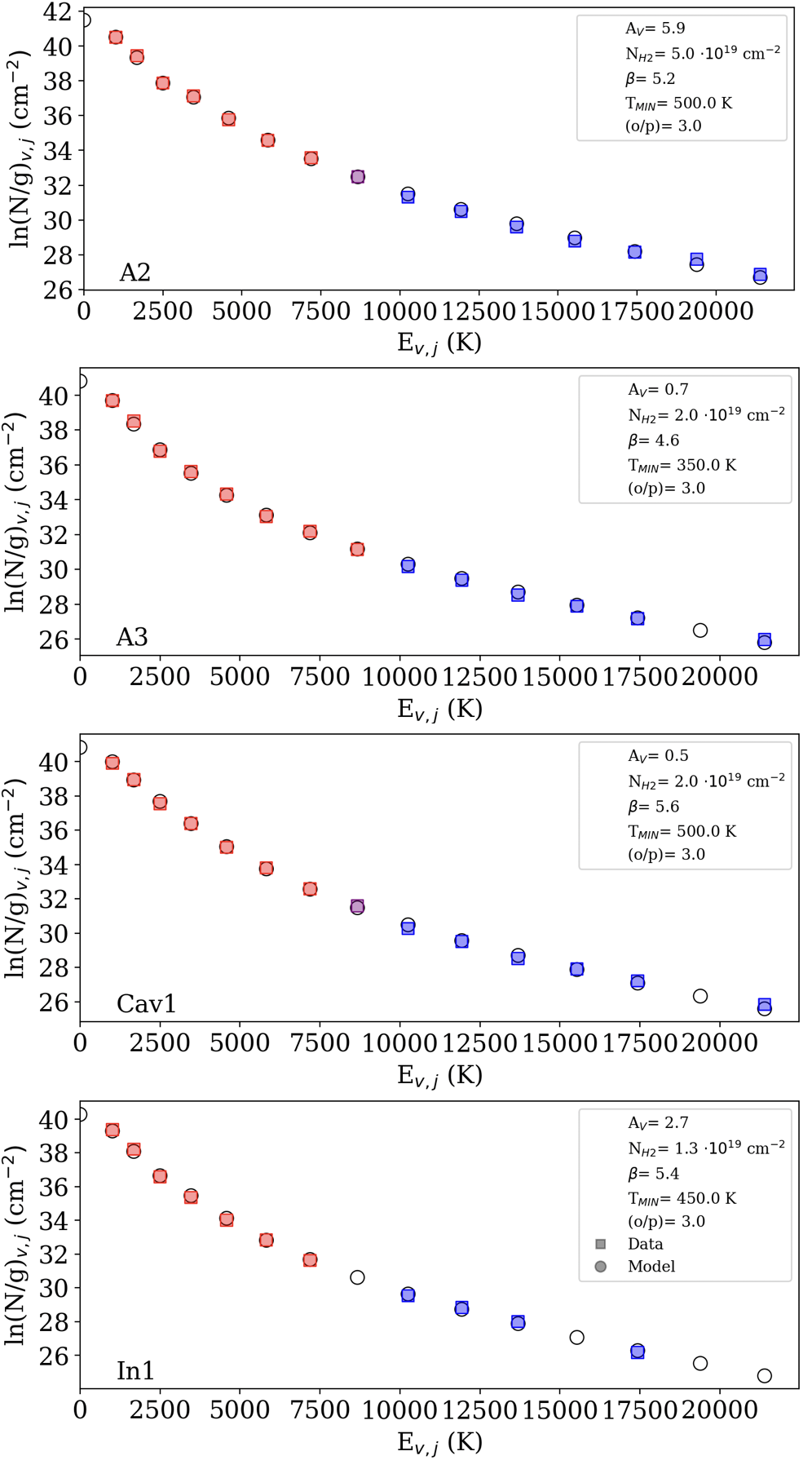}
    \caption{Rotational diagrams for pure rotational v=0 lines. Squares represent the data. Colors correspond to the different instruments, with red and blue being MIRI and NIRSpec respectively. The black empty circles are the best model for each region. }
    \label{fig:temp_str_reg}
\end{figure}

\begin{figure*}[ht]
    \centering
    \includegraphics[width=1\textwidth,keepaspectratio]{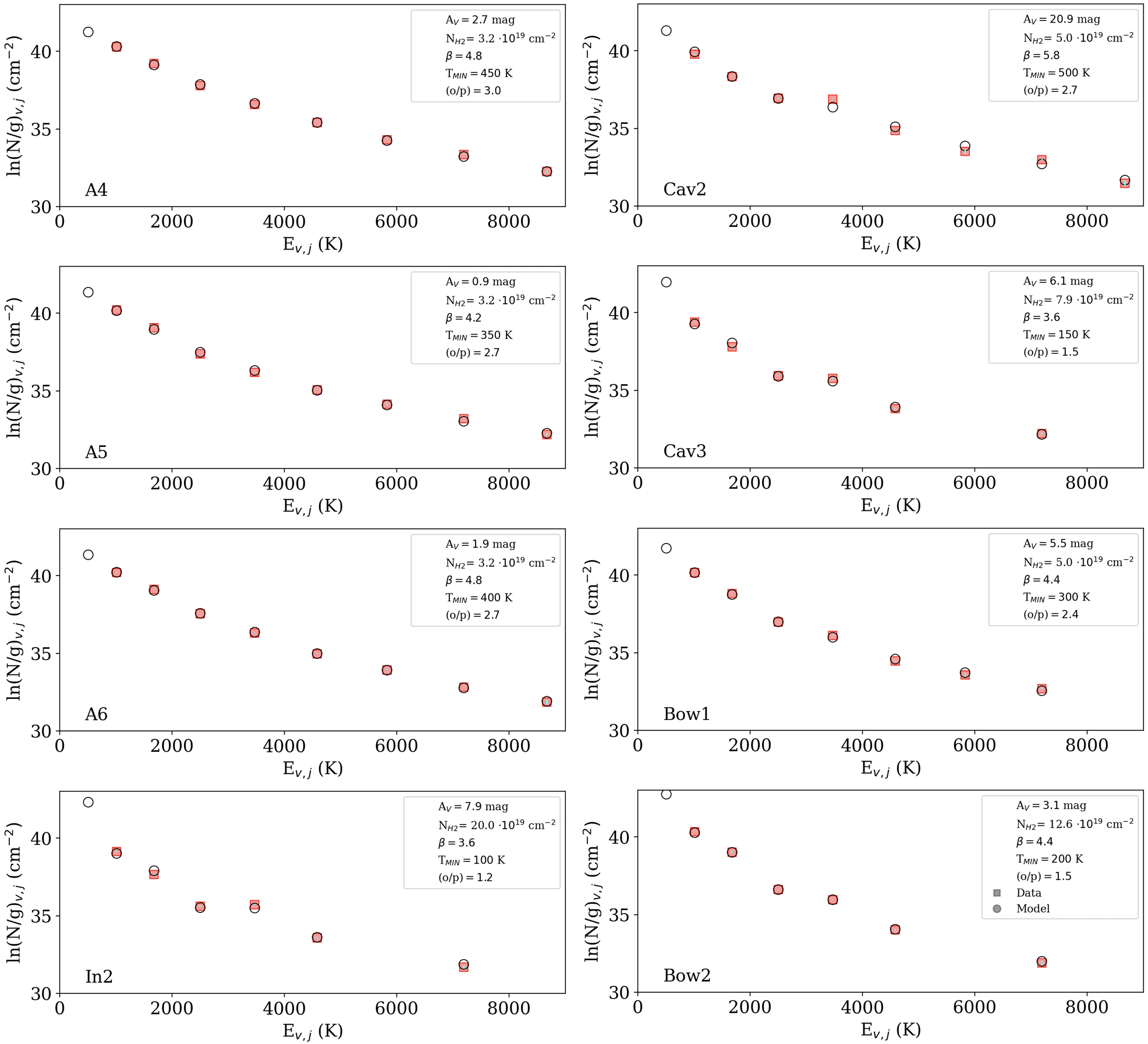}
    \caption{Rotational diagrams for pure rotational v=0 lines obtained for regionswhere just MIRI data is available. Squares represent the data. The black empty circles are the best model for each region. }    \label{fig:ap_temp_str_MIRIALL}
\end{figure*}

This section presents the results obtained from modeling the rotational diagrams assuming a temperature stratification. Figure \ref{fig:temp_str_reg} displays the rotational diagrams for pure rotational v=0 lines in regions observed with both MIRI and NIRSpec, along with the best-fitting stratification model and the derived parameters for each case. Figure \ref{fig:ap_temp_str_MIRIALL} shows the results for regions observed only with MIRI.

\clearpage

\section{Kinematic parameters and velocity maps for pure rotational lines}
\begin{figure}[h]
    \centering
\includegraphics[width=0.8\textwidth,keepaspectratio]{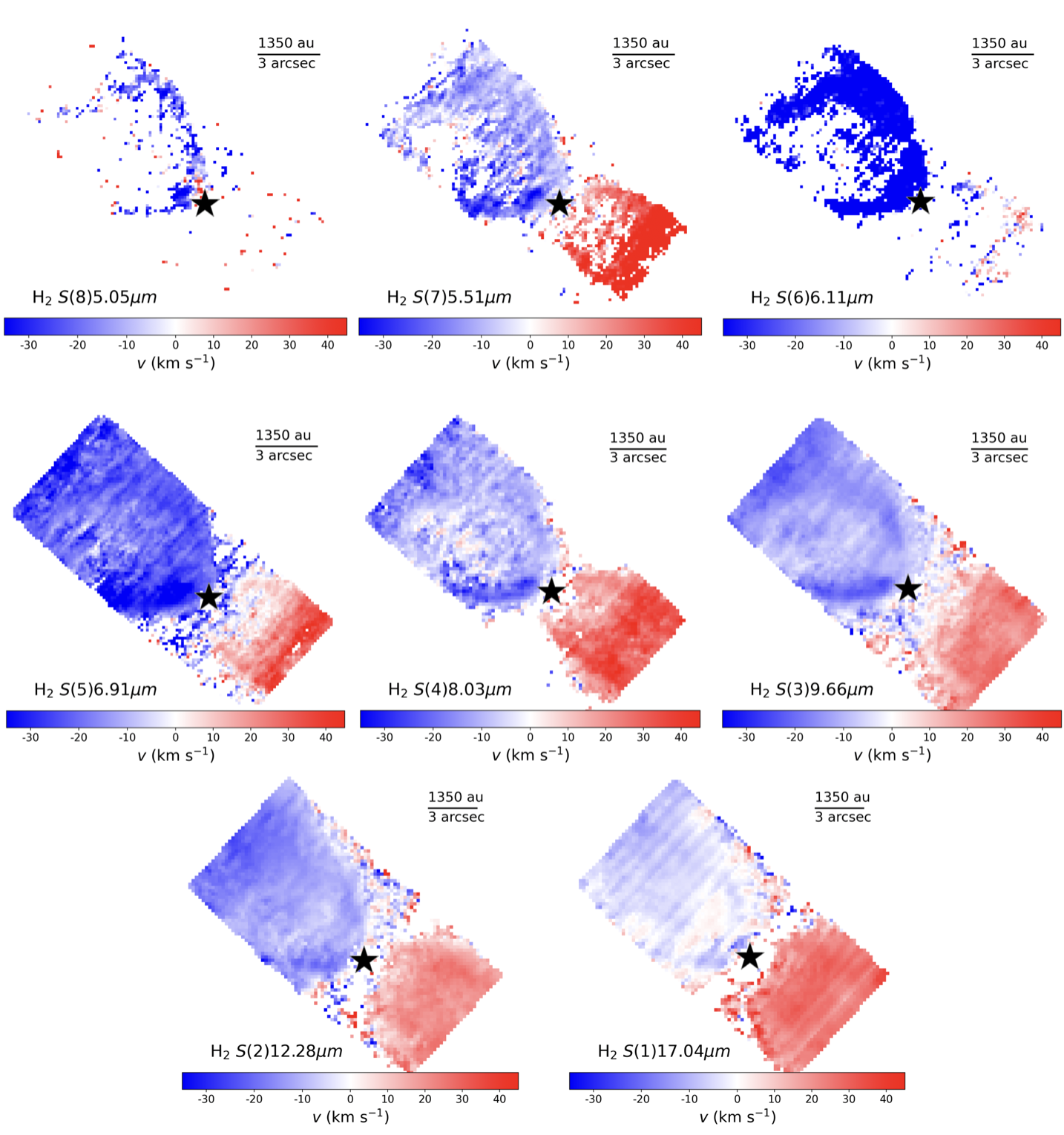}
\caption{
H$_2$ radial velocity maps obtained from the pure-rotational 0--0~S(1) to S(8) transitions. 
Black stars indicate the source positions. 
The MIRI FoV varies with wavelength, from 6\arcsec$\times$15\arcsec\ for the 0--0~S(8) map at 5.05~\um\ to 8\arcsec$\times$17\arcsec\ for the 0--0~S(1) map at 17.04~\um. 
The overall morphology remains similar across lines tracing different excitation conditions, although the mean velocity values vary inversely with wavelength. 
The same color scale is adopted for all maps to enable direct comparison.}
    \label{fig:vel_maps_miri}
\end{figure}

This section details the results of the kinematic study of HH46. Figure \ref{fig:vel_maps_miri} presents the \htwo\ radial velocity maps obtained with MIRI for the pure rotational lines S(1) to S(8), complementing Figure \ref{fig:vel_miri8}. The morphology of these maps remains consistent across transitions, while the absolute velocity values vary. To highlight the excitation-dependent variation in velocity, the same colorbar limits were used across all maps. Table \ref{tab:kinematics} provides the peak velocities and FWHM values of selected transitions with different excitation energies across the various outflow regions.

\clearpage
\section{Comparison with shock models in the representative regions}

\begin{figure}[ht]
    \centering
\includegraphics[width=1\textwidth,keepaspectratio]{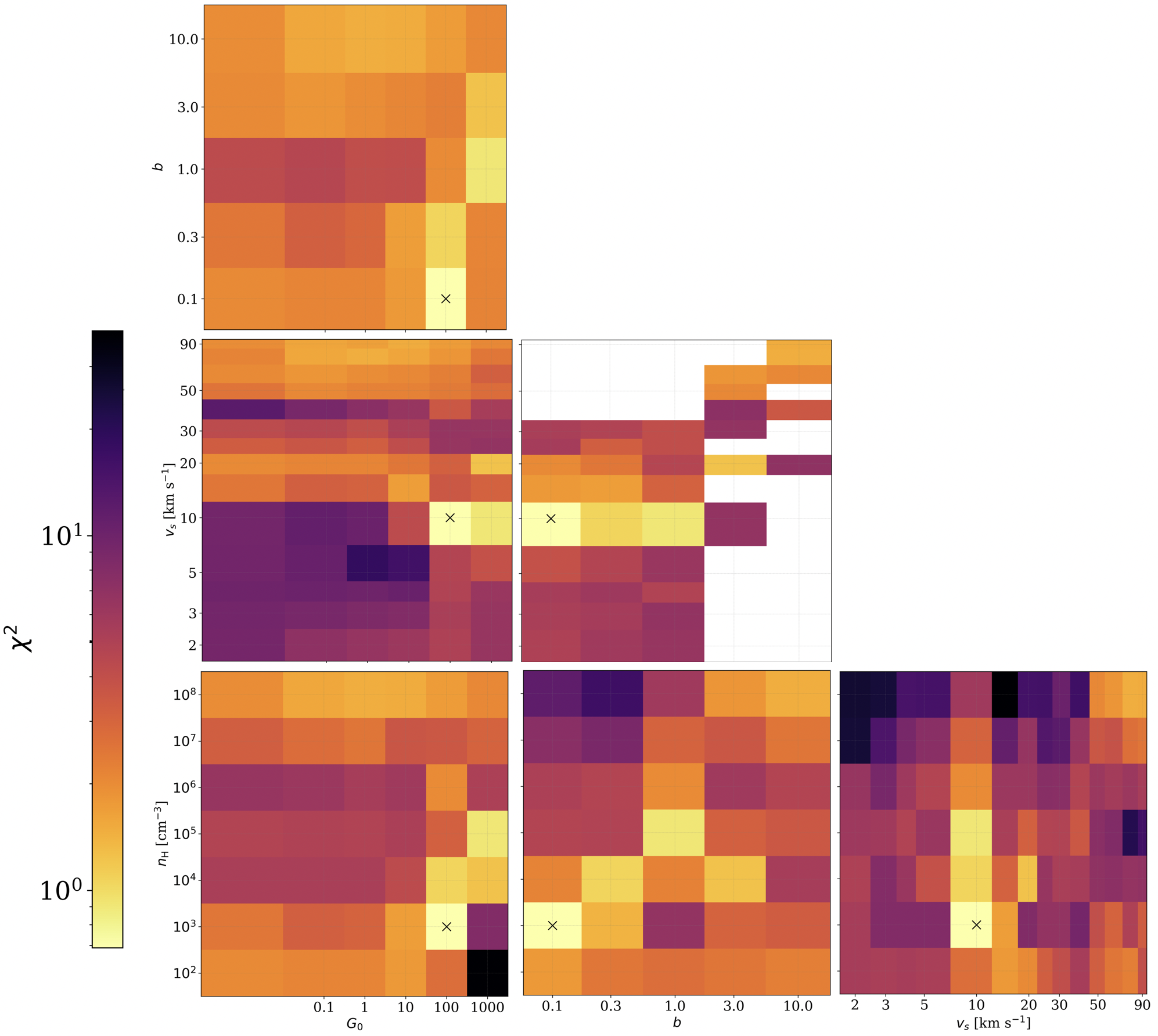}
    \caption{Corner plot showing the $\chi^2$ distribution across the model parameter space for knot A1. Each panel represents the $\chi^2$ minimization map for a pair of free parameters ($b$, $v_s$, $n_{\rm H}$, and $G_0$), with $\chi^2$ values minimized over all remaining parameters. Lighter colors correspond to lower $\chi^2$ values, indicating better agreement between the observed and modeled H$_2$ column densities. The best-fit solution is marked with a black cross.
    }
    \label{fig:chisq}
\end{figure}

Figure~\ref{fig:chisq} presents the $\chi^2$ corner plot for knot A1, showing how the $\chi^2$ values vary across the explored parameter space and are minimized for each pair of free parameters. The parameters considered are the shock velocity ($v_s$), UV radiation field strength ($G_0$), transverse magnetic field strength ($b$), and preshock density ($n_H$), while the cosmic-ray ionization rate and PAH abundance were fixed at $\zeta_{H_2}=10^{-17}$~s$^{-1}$ and $\chi$(PAH)=$10^{-8}$, respectively. This visualization highlights parameter correlations and provides a quantitative assessment of the robustness of the best-fit solution.

\begin{figure}[ht]
    \centering
\includegraphics[width=0.5\textwidth,keepaspectratio]{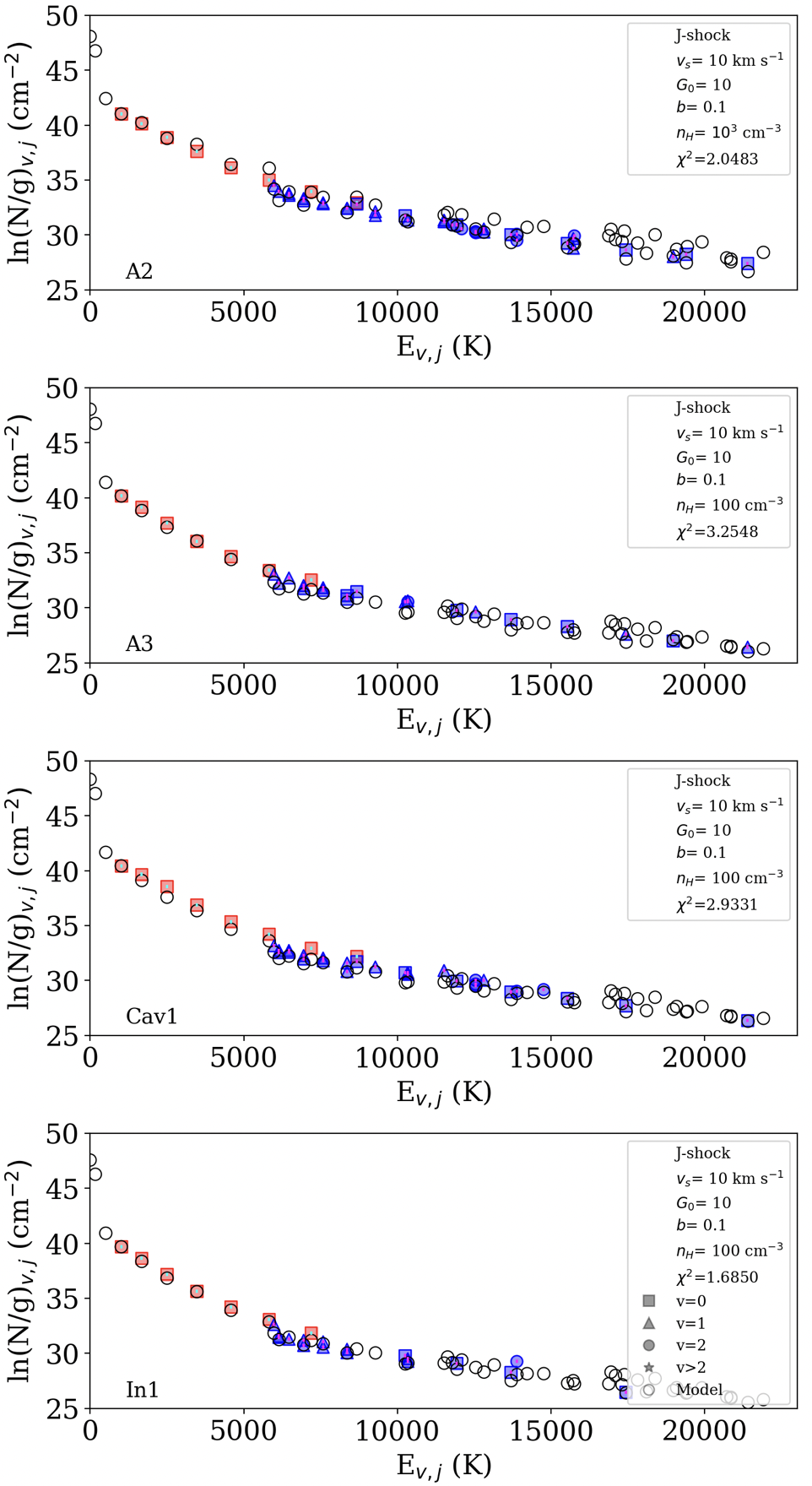}
    \caption{Best shock models (black open circles) for the representative regions. Red and blue points represent MIRI and NIRSpec data, respectively, while the symbols —squares, triangles, circles, and stars—correspond to vibrational levels from v=0 to v$>$2. }
    \label{fig:shockmodall}
\end{figure}

Figure~\ref{fig:shockmodall} complements Figure~\ref{fig:shock_modelA1} by showing the best-fit shock models for all other regions observed with MIRI and NIRSpec. The corresponding fitted parameters are shown as insets in each panel and are summarized in Table~\ref{tab:shockmod}. 

 \begin{table} 
 \centering 
 \caption{Parameters obtained for the best shock models, including $v_s$: shock velocity, $G_0$:UV radiation field strength, $b$: transverse magnetic field strength and $n_H$: pre-shock density. The cosmic-ray ionization rate ($\zeta_{H2}$) and the abundance of polycyclic aromatic hydrocarbons (PAHs, $X$(PAH)) where fixed to the minimum possible value of $\zeta_{H2}= 10^{-17} s^{-1}$ and $\chi$(PAH)= $10^{-8}$, respectively.} 
 \label{tab:shockmod} 
\begin{tabular}{ c c c c c c c c } 
 \hline\hline 
   \multicolumn{1}{c|}{ }   & \multicolumn{1}{c|}{Type }  & \multicolumn{1}{c|}{$v_s$ }  & \multicolumn{1}{c|}{$G_0$ }  & \multicolumn{1}{c|}{$b$ }  & \multicolumn{1}{c|}{$n_H$ } \\  
   
   \multicolumn{1}{c|}{ Reg}   & \multicolumn{1}{c|}{ }  & \multicolumn{1}{c|}{\tiny ($km s^{-1}$ )}  & \multicolumn{1}{c|}{}  & \multicolumn{1}{c|}{}  & \multicolumn{1}{c|}{\tiny ($ cm^{-3}$ )} \\  
 \hline 
 
A1 		& J-Shock 	& 10	& 100 	& 0.1 	& $10^3$   \\ 
A2 		& J-Shock 	& 10 	& 10 	& 0.1 	& $10^3$  \\ 
A3 		& J-Shock 	& 10 	& 10 	& 0.1 	& $10^2$ \\ 
Cav1 	& J-Shock	& 10 	& 10 	& 0.1 	& $10^2$ \\ 
In1 	& J-Shock 	& 10 	& 10 	& 0.1 	& $10^2$ \\ 
  
 \hline 
 \hline 
\end{tabular} 
 \end{table}

\bibliographystyle{aasjournal}



\end{document}